%% file: sample-manuscript.tex
\documentclass[acmsmall,screen]{acmart}

\AtBeginDocument{%
  }

\setcopyright{acmlicensed}
\copyrightyear{2018}
\acmYear{2018}
\acmDOI{XXXXXXX.XXXXXXX}
\acmConference[Conference acronym 'XX]{Make sure to enter the correct
  conference title from your rights confirmation email}{June 03--05,
  2018}{Woodstock, NY}
\acmISBN{978-1-4503-XXXX-X/2018/06}

\usepackage{listings}
\usepackage{caption}
\usepackage[most]{tcolorbox}
\usepackage{lipsum}
\tcbuselibrary{listingsutf8}
\usepackage{multirow}
\usepackage{algorithm}
\usepackage{algpseudocode}
\usepackage{makecell}
\usepackage[commandnameprefix=ifneeded, final]{changes} %最终版
\setdeletedmarkup{} %使用这一行不显示deleted内容

\begin{document}

\algnewcommand\algorithmicinput{\textbf{Input:}}
\algnewcommand\Input{\item[\algorithmicinput]}
\algnewcommand\algorithmicoutput{\textbf{Output:}}
\algnewcommand\Output{\item[\algorithmicoutput]}
\algtext*{EndIf}
\algtext*{EndFor}
\algtext*{EndWhile}
%\algtext*{EndFunction}

\title{\deleted{Can Large Language Models Solve Path Constraints in Symbolic Execution?}\added{Can Large Language Models Reason About Complex Execution Paths? An Empirical Study on Python}}

\author{Wenhan Wang}
\authornote{Both authors contributed equally to this research.}
\email{wangwenhan@isacas.ac.cn}
\affiliation{%
  \institution{Institute for Software Chinese Academy of Sciences}
  \country{China}
}

\author{Kaibo Liu}
\authornotemark[1]
\email{liukb@pku.edu.cn}
\affiliation{%
  \institution{Peking University}
  \country{China}
}

\author{Zeyu Sun}
\email{zeyu.zys@gmail.com}
\affiliation{%
  \institution{Institute for Software Chinese Academy of Sciences}
  \country{China}
}

\author{An Ran Chen}
\email{anran6@ualberta.ca}
\affiliation{%
  \institution{University of Alberta}
  \country{Canada}
}

\author{Ge Li}
\email{lige@pku.edu.cn}
\affiliation{%
  \institution{Peking University}
  \country{China}
}

\author{Gang Huang}
\email{hg@pku.edu.cn}
\affiliation{%
  \institution{Peking University}
  \country{China}
}

\author{Lei Ma}
\email{ma.lei@acm.org}
\affiliation{%
  \institution{The University of Tokyo / University of Alberta}
  \country{Japan / Canada}
}

\renewcommand{\shortauthors}{Anonymous et al.}

\begin{abstract}
%\replaced{新内容}{旧内容}
\deleted{Symbolic execution is a fundamental software analysis technique that supports tasks such as testing and debugging, yet its real-world applicability remains limited. A key obstacle is its difficulty in solving diverse path constraints: SMT-based approaches struggle with complex data structures and external API calls.}

\added{Execution path reasoning is a key step towards program semantics understanding. It is crucial for generating test cases that cover certain branches/paths, or detecting bugs that are triggered by some paths without actually executing the program. Traditionally, execution path reasoning can be achieved by symbolic execution techniques, but existing SMT-based symbolic execution approaches struggle with complex data structures and external API calls.}
This challenge is even more pronounced in languages with highly flexible syntax, such as Python, resulting in a lack of widely adopted tools for reasoning on execution paths.
\added{Therefore, reasoning execution paths with AI-based approaches become a promising direction.}

In this paper, \deleted{we investigate the feasibility of adopting large language models (LLMs) for solving path constraints when traditional symbolic execution tools cannot be applied, e.g., Python. We conduct an empirical study to evaluate the ability of LLMs in two types of path constraint solving: generation tasks for test case generation and classification tasks for bug detection. We build new evaluation pipelines and benchmarks from both competition-level programs and real-world repositories. Our results show that state-of-the-art LLMs can solve path constraints and improve test coverage on real-world software, though models with stronger reasoning abilities do not always outperform weaker ones. These findings highlight the possibility of extending symbolic execution techniques with LLMs to improve the generalizability of symbolic execution.}
\added{we investigate the feasibility of adopting large language models (LLMs) for execution path reasoning on Python, where traditional path-based symbolic execution tools are unavailable. We conduct an empirical study on two types of path reasoning tasks: generation tasks for test case generation and classification tasks for bug detection. We build new evaluation pipelines and benchmarks from both competition-level programs and real-world repositories. Our results show that state-of-the-art LLMs can perform correct reasoning on execution paths and improve test coverage on real-world software, though models with stronger reasoning abilities do not always outperform weaker ones. These findings highlight the potential of utilizing LLMs as a complementary heuristic for path-aware code reasoning, especially in program languages lacking mature symbolic execution tools. We have released our benchmark and evaluation scripts at \url{https://github.com/jacobwwh/llm-path-study}.}

%Our experiment results show that state-of-the-art LLMs are able to solve path constraints in both generation and classification tasks, with 60\% of generated test cases successfully covering their intended execution paths. Furthermore, LLMs are capable of improving test coverage by covering execution paths in real-world repositories where traditional symbolic execution tools cannot be applied. These findings highlight the possibility of extending symbolic execution techniques with LLMs in the future to improve the ability and generalizability of symbolic execution.
\end{abstract}


%\ccsdesc[500]{Do Not Use This Code~Generate the Correct Terms for Your Paper}
%\ccsdesc[300]{Do Not Use This Code~Generate the Correct Terms for Your Paper}
%\ccsdesc{Do Not Use This Code~Generate the Correct Terms for Your Paper}
%\ccsdesc[100]{Do Not Use This Code~Generate the Correct Terms for Your Paper}


\received{20 February 2007}
\received[revised]{12 March 2009}
\received[accepted]{5 June 2009}

\maketitle

\section{Introduction}

\input{sections/introduction}

\section{Background and Related Work}
\input{sections/related}

\section{Methodology}
\input{sections/approach}

\section{Study Results}
\input{sections/evaluation}

\section{Discussion}
\input{sections/discussion}

\section{Threats to Validity}
\input{sections/threats}

\section{Conclusion}

In this paper, we explore the capability of large language models in \deleted{solving} \added{reasoning on} complex execution paths. 
%which \deleted{is the cornerstone of symbolic execution} \added{resembles the symbolic execution technique where SMT solvers are adopted to solve path constraints}. 
%We conduct a systematic study on Python,
%\added{on Python, which is challenging for symbolic execution tools,} 
We focus on two key applications of path reasoning: test case generation and execution path classification for bug detection. Our evaluation spans diverse program domains and challenging scenarios, ranging from complex algorithms in programming contests to large-scale real-world repositories. The results demonstrate that state-of-the-art LLMs possess strong potential for understanding and reasoning about execution paths, with reasoning and non-reasoning models excelling in different tasks. Nevertheless, several challenges remain, including difficulties in interpreting API behaviors and performance degradation caused by overthinking in LRMs. Overall, our findings provide evidence that LLM-powered \deleted{or LLM-augmented symbolic execution} \added{execution path reasoning can act as a substitution for symbolic execution to a certain extent,} \deleted{is feasible} and highlight promising directions for future enhancements\deleted{on LLM-based path constraint solving}. 
\added{In the future, we aim to extend our study to other programming languages and symbolic execution techniques, and improve the ability of LLMs in execution path reasoning via post-training or in-depth integration with symbolic analysis tools.}

%\begin{acks}
%\end{acks}

%\section*{Data Availability}
%The code of our study is available at \url{https://anonymous.4open.science/r/llm4sym-study-767C}.

\bibliographystyle{ACM-Reference-Format}
\bibliography{sample-base}

%\appendix

\end{document}

%% file: sections/introduction.tex
\deleted{Symbolic execution \cite{king1976symbolic, baldoni2018survey} is an important program analysis technique that supports a wide range of software engineering applications, such as testing \cite{cadar2008klee}, bug finding \cite{ramos2015under}, and verification \cite{coen2001using}. The key idea of symbolic execution is to execute the program with symbolic inputs instead of concrete values, and simultaneously explore different execution paths. During this process, symbolic execution first traverses the program to collect the path condition constraints and subsequently employs an SMT solver \cite{de2008z3} (e.g., Z3 or CVC5) to determine their satisfiability and predict possible values for symbolic variables.}                        
                                                  
\added{Execution path reasoning is a fundamental capability that underpins many software engineering tasks, including testing \cite{cadar2008klee}, bug finding \cite{ramos2015under}, and verification \cite{coen2001using}. Traditionally, symbolic execution \cite{king1976symbolic, baldoni2018survey} has been the primary technique for this purpose: it executes the program with symbolic inputs instead of concrete values, traverses different execution paths to collect path condition constraints, and subsequently employs an SMT solver \cite{de2008z3} (e.g., Z3 or CVC5) to determine their satisfiability and predict possible values for symbolic variables.}

\deleted{Although symbolic execution has become a mature technique for software engineering, several critical challenges still hinder its broader adoption in general-purpose program analysis. Among these challenges, one of the most fundamental lies in solving complex path constraints. First, for some program languages (especially object-oriented program languages), path constraints with complex data structures are difficult to transfer into SMT-compatible constraints. Moreover, the complexity of program dependencies in real-world software repositories, such as the wide usage of external APIs, also makes existing symbolic execution tools not applicable. The above limitations of symbolic execution prevents this technique from further application, and urges for substitutional approaches for analysis on execution path constraints.}

\added{Although symbolic execution has become a mature technique, several critical challenges still hinder its broader adoption as a general-purpose approach for execution path reasoning. Among these challenges, one of the most fundamental challenges lies in solving complex path constraints. First, for some program languages (especially object-oriented program languages), path constraints with complex data structures are difficult to transfer into SMT-compatible constraints. Moreover, the complexity of program dependencies in real-world software repositories, such as the wide usage of external APIs, also makes existing symbolic execution tools not applicable. These limitations mean that, in practice, execution path reasoning remains an unsolved problem for many languages and application scenarios, and calls for alternative approaches that can reason about execution paths without relying on traditional SMT-based constraint solving.}

Recently, large language models (LLMs) have demonstrated remarkable capabilities in both code generation and code understanding \cite{jimenez2024swebench, gu2024cruxeval}, achieving performance comparable to, or even surpassing, that of human experts across a wide range of tasks. More significantly, LLMs have begun to reshape areas of software engineering that were once dominated by symbolic execution, including automated testing \cite{wang2024software, wang2025testeval, jain2025testgeneval} and bug detection \cite{li2024enhancing}. These successes suggest that LLMs possess a latent understanding of program semantics and execution paths, \deleted{suggesting the potential to complement or enhance symbolic execution techniques} \added{making them a promising alternative for execution path reasoning where traditional symbolic execution falls short}. The recent emergence of large reasoning models (LRM) \cite{li2025system} further strengthens this potential. By integrating reinforcement learning, LRMs significantly strengthen the reasoning abilities of LLMs, enabling them to generate extended chains of thought (CoT) \cite{wei2022chain} before producing final answers. These models excel at complex logical reasoning tasks, such as mathematical problem solving and theorem proving, that \deleted{form the very foundation of symbolic execution. Since symbolicexecution intrinsically relies on constraint solving and systematic path exploration through logical reasoning, it appears to be a natural beneficiary of LLM-driven enhancement} \added{closely relate to the skills needed for execution path reasoning, such as constraint solving and systematic path exploration.} This convergence of needs and capabilities raises a pivotal question: Can LLMs \deleted{empower symbolic execution to advance software engineering tasks} \added{reason about complex execution paths in advancing software engineering tasks}?

In our study, we select Python as the study objective. This choice is motivated by Python's highly flexible grammar features (e.g., dynamic typing), which limit the availability of mature off-the-shelf program analysis tools \cite{li2022scalpel}. For example, there are no well-recognized symbolic execution tools for Python. The tools \cite{PyExSMT, CrossHair} that have been used in previous research studies \cite{jiang2024towards, xu2025identifying} are either deprecated \cite{PyExSMT} or only maintained by independent developers \cite{CrossHair}. In contrast, LLMs have demonstrated stronger reasoning abilities on Python compared to other programming languages \cite{zheng2023codegeex, yan2024codescope}. These observations suggest that LLMs could serve as a valuable complement to, or even a potential replacement for, traditional \deleted{program analysis techniques} \added{execution path reasoning approaches} in Python. 
%\added{We note that our study and conclusions are scoped to Python; whether the findings generalize to languages with mature solver-backed symbolic execution ecosystems (e.g., C/C++ with KLEE, Java with SPF) remains an open question for future work.}

In this paper, we conduct a systematic empirical study on the efficacy of LLMs in \deleted{solving path constraints for symbolic execution} \added{reasoning about execution paths on Python}, and analyze the different behaviors of non-reasoning LLMs and LRMs. We focus on evaluating LLMs on two different tasks, which include both generation-style and classification-style. The first one is generating test cases to satisfy a given execution path, which is a typical application of symbolic execution. The second task is execution path classification: determine whether a path is valid or if it can trigger a bug. To evaluate these tasks, we construct benchmarks derived from both programming competition problems—characterized by intricate control flows—and real-world software repositories, which often involve the use of third-party libraries. We aim to address the following research questions:

\begin{itemize}
      \item \textbf{RQ1: Can LLMs generate test cases \deleted{for path
  constraints} \added{that cover given execution paths}?} We investigate the power of LLMs in test case generation, \added{a key downstream task for execution path understanding, and a standard application field for symbolic execution}. For this task, we evaluate LLMs by prompting them to generate test cases for Python programs to cover given execution paths. Although prior work \cite{ryan2024code} has attempted to enhance LLM-based test case generation with execution path information, it lacks support for paths with complex control flow structures, such as loops, and it remains unexplored whether the generated test case can precisely cover the given path.

  \item \textbf{RQ2: Can LLMs categorize \deleted{path constraints} \added{execution paths} by their feasibility and bug-triggering behaviors?} \deleted{Symbolic execution is also widely used in classification-style tasks, such as classifying path feasibility for bug detection.} \added{Path classification, such as path feasibility analysis, is also an important topic for program analysis. In this RQ, we study the ability of LLMs in execution path reasoning for supporting classification-style tasks.} \deleted{For this RQ, we} \added{We} extract both feasible and infeasible execution paths by designing a path traversal algorithm on 
  control flow graphs (CFGs). Based on these extracted paths, we then evaluate whether LLMs can accurately discriminate among valid, infeasible, and bug-triggering execution paths.                

  \item \textbf{RQ3: Can LLMs handle \deleted{path constraint solving} \added{execution path reasoning} in real-world software repositories?} In the previous two RQs, we mainly focused on \deleted{path constraint solving} \added{execution path reasoning} in standalone programs without external dependencies, which may not reflect real-world software development practices. \deleted{Among the key challenges of symbolic execution, external} \added{In contrast to LLMs, external} API calls represent a major obstacle that hinders the applicability of traditional symbolic execution to practical software systems. For this RQ, we investigate whether LLM-based \deleted{path constraint solving} \added{execution path reasoning} can help software testing on real-world software repositories, i.e., if prompting LLMs to cover designated execution paths can improve test coverage.
  \end{itemize}
               
Our study shows several findings that highlight the ability of LLMs and provide insights for future research. First, state-of-the-art LLMs, especially LRMs, are capable of generating correct test inputs that satisfy execution paths with complex control flows. For example, state-of-the-art LLMs generate \added{correct test cases for} more than 65\% of the execution paths in our challenging benchmark. Second, LLMs can, to a certain extent, distinguish feasible execution paths from those that are infeasible or bug-triggering. However, the extensive reasoning process of LRMs may occasionally mislead them into erroneous predictions. Third, LLM-based \deleted{path constraint solving} \added{execution path reasoning} can help software testing in real-world repositories by improving test coverage, \added{while the test pass rate still remains a bottleneck}. These findings point to the broader applicability of LLMs in \deleted{path constraint analysis and symbolic execution} \added{execution path reasoning, especially for languages lacking powerful symbolic execution tools}.

In summary, the main contributions of this paper are:

\begin{itemize}
    \item To our knowledge, our work presents the first systematic empirical study on the capability of LLMs to \deleted{solve path constraints for symbolic execution} \added{reason on execution paths constraints for the dynamically-typed language Python}.

    \item We construct new benchmarks that enable in-depth evaluation of execution path analysis tasks. These benchmarks encompass a broad spectrum of downstream applications\deleted{of symbolic execution}, including test case generation, path feasibility analysis, and bug detection.

    \item We uncover both the strengths and limitations of LLMs (reasoning and non-reasoning) in addressing complex path constraints across generation and classification tasks. These findings shed light on the potential of LLMs to enhance or replace traditional symbolic execution \added{for execution path reasoning}.
\end{itemize}

%% file: sections/related.tex
\subsection{Large Language Models for Program Execution Reasoning}

Recently, there has been a trend in understanding the ability of LLMs in code-related tasks beyond code generation. An important direction is code execution reasoning, where LLMs are required to simulate program execution or answer questions about execution behavior. CRUXEval \cite{gu2024cruxeval} is the first comprehensive benchmark for LLM-based code execution reasoning, encompassing two basic tasks: input prediction and output prediction. A similar benchmark, REval \cite{chen2025reasoning}, further designed new evaluation tasks such as fine-grained code coverage prediction and program state prediction. PathEval \cite{jiang2024towards} evaluated both LLMs and symbolic execution tools for predicted test inputs given target program outputs.
Ni et al. \cite{ni2024next} proposed NEXT, an approach that enhances code execution reasoning by fine-tuning LLMs on program execution trace data. SemCoder \cite{ding2024semcoder} trains a code LLM with operational semantics related to code execution, resulting in notable improvements on downstream tasks such as code generation and execution reasoning. Armengol et al. \cite{armengol2025cannot} further proposed a ``dynamic scratchpad'' mechanism, which augments LLMs with external memory to track variable states during execution. 
%The dynamic scratchpad improves both general-purpose LLMs and execution-tuned LLMs on CRUXEval.

\subsection{LLM-integrated Symbolic Execution}

%Symbolic execution operates on path constraints. For a program $S$, symbolic execution first extracts execution path constraints $P$, which can be represented by:

%\begin{equation}
    %P=\left\{ (C_{1}, B_{1}), (C_{2}, B_{2}), ..., (C_{n}, B_{n})\right\}
%\end{equation}

%Where $C_{i}$ denotes a code block (with one or more statements), and $B_{i}$ is the branch condition that $P$ takes at the last statement of $C_{i}$ (if $C_{i}$ does not end with a branch or loop condition, then $B_{i}$ is empty). The essence of symbolic execution is to solve $P$, which requires reasoning about whether all $C_{i}$ can be simultaneously satisfied. For test case generation, symbolic execution generates a group of valid input parameters that can satisfy $P$. For bug detection, symbolic execution checks whether satisfying $P$ will result in violating specifications related to certain bug types (e.g., division by zero).

Recently, researchers have been investigating the possibility of integrating the concept of symbolic execution into LLM-driven software testing. Symprompt \cite{ryan2024code} generates test cases by prompting LLMs with execution paths extracted from ASTs, but it does not pay special consideration to paths with loops. Wang et al. \cite{wang2024python} proposed LLM-Sym, which converts complex execution path constraints into Z3 code with an LLM agent. AutoBug \cite{li2025large} tried to adopt LLMs to reason on execution paths and slices for symbolic execution. Wu et al. \cite{wu2025generating} proposed PALM, which improves test coverage by using LLMs to solve execution paths represented in a program variant form. While these approaches demonstrate promising results, they primarily focus on relatively simple programs with only a limited number of branches (e.g., those derived from HumanEval \cite{chen2021evaluating}), leaving complex control flow structures largely unexplored. Unlike the above works that tried to replace symbolic execution tools with LLMs, Cottontail \cite{tu2026cottontail} leverages LLMs to augment an existing concolic executor to generate structured test inputs. \added{NexusSym \cite{wang2025nexusym} proposed a hybrid framework to augment the symbolic execution tool KLEE with LLMs. WARP \cite{koh2025worst} is another tool-integrated approach that produces and generalizes worst-case constraints for symbolic execution with Java PathFinder and LLMs. SESpec \cite{yang2025integrating} integrates symbolic execution and LLMs for formal specification generation.}

\added{Although these works tried to adopt LLMs for symbolic execution, they do not target a key question: ``can LLMs directly reason on complex execution paths without the aid of existing symbolic execution tools?'' AutoBug \cite{li2025large} focuses on proving postconditions rather than reasoning specific execution paths. Symprompt \cite{ryan2024code} and PALM \cite{wu2025generating} focus on generating test cases for trivial execution paths extracted from ASTs, without considering complex execution flows and non-generative tasks. NexusSym \cite{wang2025nexusym}, WARP \cite{koh2025worst}, Cottontail \cite{tu2026cottontail}, and ConcoLLMic \cite{luo2026agentic} focus on extending the capability of symbolic execution tools with LLM. While our study targets the question by conducting a comprehensive study on LLM-driven execution path reasoning across multiple tasks and different fields, with consideration of challenges brought by both complex control flow and real-world software repositories.}

%% file: sections/approach.tex
\begin{figure}[h]
  \centering
  \includegraphics[width=0.8\linewidth]{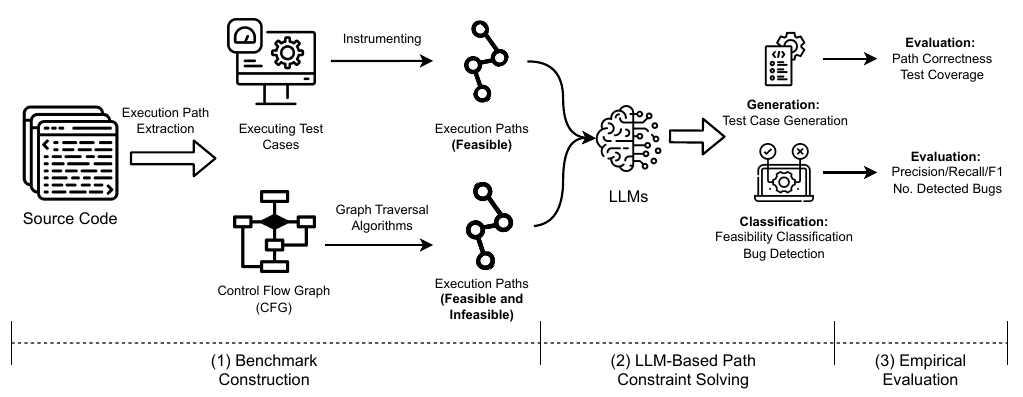}
  \caption{An overview of our empirical study.}
  \label{fig:overview}
\end{figure}

Figure~\ref{fig:overview} presents an overview of our study methodology. Our goal is to investigate the effectiveness of existing LLMs in understanding and solving execution path constraints. Our evaluation framework consists of two core tasks: test case generation and path classification. For these tasks, we begin by constructing benchmarks and extracting path constraints from competition problems to assess LLM performance under complex yet controlled settings. Moreover, we evaluate LLMs on a repository-level test case generation benchmark to analyze the capability of LLMs in addressing execution path constraints in real-world software.

%\subsection{Data Collection}

\subsection{Test Case Generation (competition-level)}

\subsubsection{Benchmark Construction}
The key question of the test case generation task is: given an execution path of a program under test, can LLM generate a test case for the program under test that accurately covers this execution path? We formalize this question to: given an execution path $P=\left\{ (C_{1}, B_{1}), (C_{2}, B_{2}), ..., (C_{n}, B_{n})\right\}$ for a program $S$, where $C_{i}$ denotes a code block and $B_{i}$ is the branch condition taken at the end of $C_{i}$, the LLM is asked to generate a test case for $S$ that reproduce $P$.

For the test case generation task, we first leverage the TestEval \cite{wang2025testeval} benchmark as our data source. TestEval is a Python test case generation benchmark that consists of 210 programs under test. These programs are collected from LeetCode problem solutions with a cyclomatic complexity greater than 10, with an average complexity of 13.35. This ensures that all TestEval programs involve non-trivial control flows, such as multiple branches embedded within nested loops. The high complexity of these programs allows us to rigorously evaluate path constraint reasoning on long execution paths with numerous branch conditions. 

Although the original TestEval benchmark includes a ``targeted path coverage'' task, it is not suitable for our study because it only requires LLMs to cover a short and incomplete execution path of merely 5 statements. Instead, we construct complete execution path data by running example test cases from the LeetCode problem descriptions on the TestEval programs. For this task, we deliberately avoid collecting paths through control flow graph (CFG) traversal, as such an approach often produces a large number of invalid execution paths, which provide little value for test case generation. Furthermore, this experiment is designed solely to evaluate path constraint solving capabilities, making it unnecessary to align strictly with real-world use of test case generation. To obtain execution paths, we modify the Python \texttt{trace} library, which records executed statements by instrumenting the running code. By default, the \texttt{trace} library only logs the executed statements themselves, which is insufficient for execution path analysis. Therefore, we extend it with the following modifications to capture richer information:

\begin{itemize}
    \item When the execution trace reaches a branch condition statement (e.g., \texttt{if}, \texttt{while}), we explicitly record whether the condition is satisfied. This is determined by inspecting the next executed statement following the branch. 
    If the subsequently executed statement is the one immediately following the branch condition, we infer that the execution path has taken the branch where the condition is true; otherwise, it corresponds to the false branch.

    \item In a \texttt{for} loop, it is crucial to keep track of the number of iterations so that LLMs can correctly infer the value of the loop variable at a given statement within the loop. 
    %To achieve this, we maintain a dictionary that records the iteration counts for all \texttt{for} loop statements during execution. 
    Whenever the execution path encounters a \texttt{for} statement, we annotate it with a short description in the form of ``Iteration: $n$,'' where $n$ starts from 1.
\end{itemize}

For each of the 210 LeetCode problems in TestEval, the problem description provides 2 to 3 example test cases. We run all these test cases to obtain example execution paths and subsequently discard any paths exceeding 1,000 statements, as such paths cannot fit within the context windows of most LLMs. After this filtering, our dataset for test case generation contains a total of 509 execution paths. On average, each execution path consists of \textbf{233 statements} and includes \textbf{103 branch statements} (covering \texttt{if}, \texttt{while}, and \texttt{for}).

\subsubsection{\deleted{Evluation} \added{Evaluation} Pipeline}
\begin{figure}[h]
  \centering
  \includegraphics[width=\linewidth]{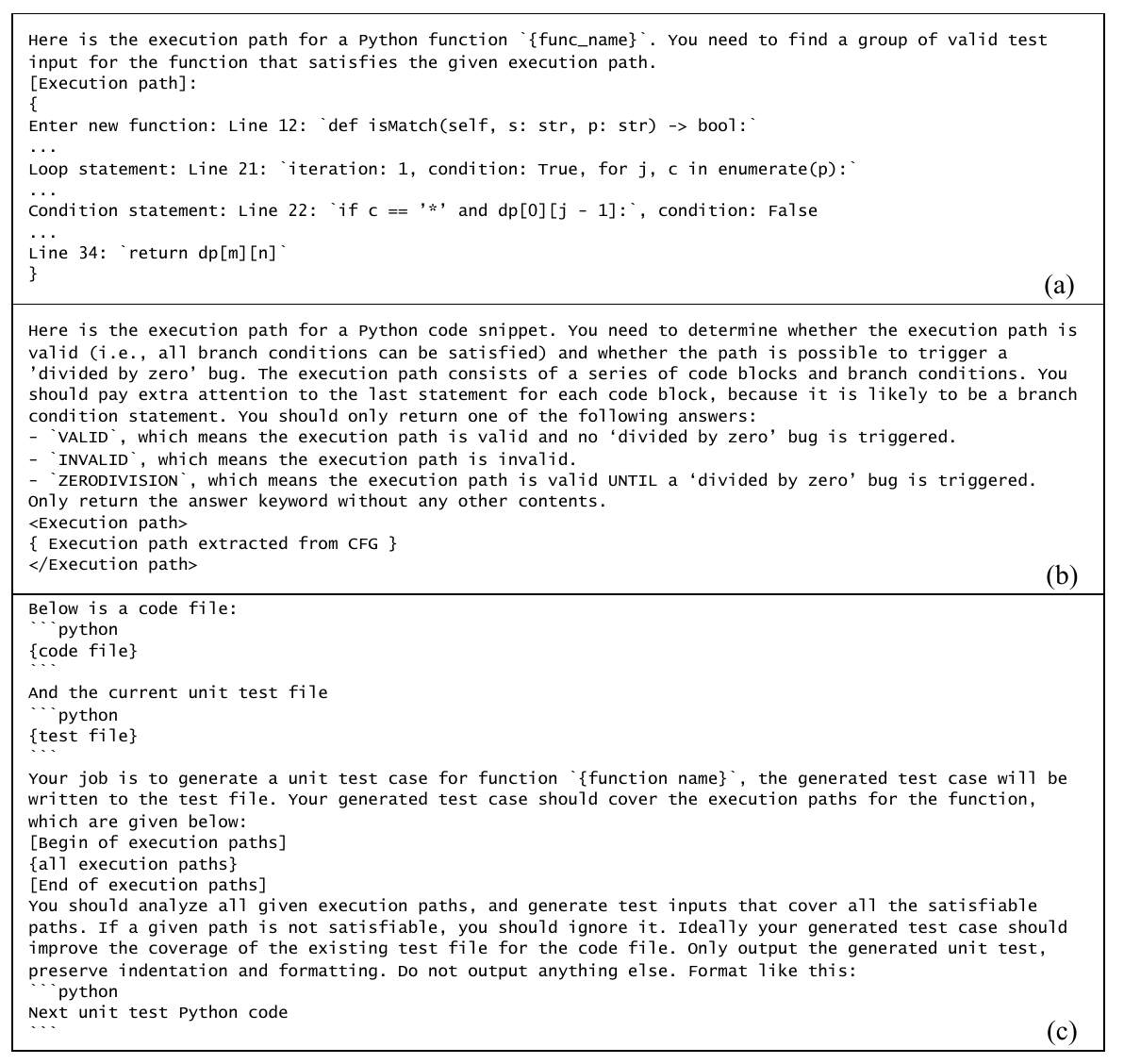}
  \caption{The prompt templates for our tasks. (a): Test case generation (competition-level) with a short example on the format of the execution path. (b): Path classification. (c): Test case generation (real-world).}
  \label{fig:template}
\end{figure}

For the competition-level test case generation task, we first create prompts for LLMs by translating the generated execution path into a more readable format. For example, we add a ``Condition statement" tag for all branch conditions in the path, and add an ``Enter function'' tag when the execution path enters a self-defined function. Figure~\ref{fig:template} (a) shows an example of the prompt for the competition-level test case generation task.

\iffalse
\begin{figure}[htbp]
\centering
\begin{tcolorbox}[listing only, listing options={basicstyle=\ttfamily}, colback=white, colframe=black,
                  width=\linewidth, sharp corners, boxrule=0.8pt,
                  fonttitle=\bfseries, fontupper=\ttfamily\small]
Here is the execution path for a Python function `{{ func\_name }}`. You need to find a group of valid test input for the function that satisfies the given execution path.

[Execution path]:

\{\{ 

Enter new function: Line 12: `def isMatch(self, s: str, p: str) -> bool:`

...

Loop statement: Line 21: `iteration: 1, condition: True,     for j, c in enumerate(p):`

...

Condition statement: Line 22: `if c == '*' and dp[0][j - 1]:`, condition: False

...

Line 34: `return dp[m][n]`

\}\}

\end{tcolorbox}
\caption{An Example of the prompt for the competition-level test case generation task.}
\label{fig:template-test}
\end{figure}
\fi

In the experiments, we observe that for certain LLMs, the prompt template shown in Figure \ref{fig:template} (a) does not always guarantee the generation of valid test inputs. Instead, these models occasionally produce descriptions of possible input parameter values in natural language or in mixed language–code forms. To address this issue, we introduce an additional post-processing step to reformat the outputs of LLMs. \deleted{Specifically, we prompt a GPT-4.1-nano LLM to transform the raw output of the path constraint–solving LLM into a correctly formatted test case.}\added{We implement a rule-based processor to extract variable values from the raw output, and try to create a test case with these variables. Then, if the rule-based processor fails or cannot extract value for certain test input arguments, we call a lightweight LLM (GPT-5.4-nano) to generate a test case from the raw output.}
Finally, we execute the generated test case on the program under test and record its execution path using our customized \texttt{trace} module. We then compare the generated execution path against the ground-truth execution path of the example test case. To evaluate the correctness of the generated test case, we design two metrics:

\begin{enumerate}
    \item \textbf{Path accuracy}: Given a generated test case with its execution path $P'=\{ (C'_{1}, B'_{1}), (C'_{2}, B'_{2}),\\ \dots, (C'_{n'}, B'_{n'}) \}$, if $P'$ is identical to the ground-truth execution path $P$, we regard the LLM as having successfully solved the execution path constraint, and assign a path accuracy score of 1 to this sample. Otherwise, the path accuracy score is 0.

    \item \textbf{Node accuracy}: In long execution paths containing dozens of branch conditions, an LLM may fail to satisfy all constraints but still correctly reason about a subset of them. This often yields a test case whose execution path partially coincides with the ground-truth path. Given a generated execution path $P'=\{ (C'_{1}, B'_{1}), \dots, (C'_{n'}, B'_{n'})\}$ and its corresponding ground-truth path $P=\{ (C_{1}, B_{1}), \dots, (C_{n}, B_{n})\}$, if they share a common prefix, i.e., $\{ (C'_{1}, B'_{1}), \dots, (C'_{k}, B'_{k})\} = \{ (C_{1}, B_{1}), \dots, (C_{k}, B_{k})\}$, then their node accuracy is defined as $Acc_{N} = \frac{k}{n}$.

\end{enumerate}

In our study, we report node accuracy at two levels of granularity: the statement level (where $C_{i}$ corresponds to a single statement) and the basic block level (where $C_{i}$ denotes a code block that terminates with a branch statement or program termination).

\subsection{Path Classification}

\subsubsection{Benchmark Construction}

For the path classification task, we construct our evaluation dataset using the Google runtime error dataset \cite{bieber2023static}. This dataset is derived from the large-scale program classification dataset CodeNet \cite{puri2codenet}, which includes code submissions from multiple online programming platforms. The Google runtime error dataset filters submissions that exhibit runtime errors and labels their error types by executing corresponding test cases. In our study, we focus specifically on programs with the error type of division by zero. 
%To ensure the difficulty of this dataset, we manually inspected the programs and removed programs that explicitly induce the bug (e.g., snippets containing \texttt{x/0} or \texttt{x\%0}). 
Additionally, we pruned programs without any control flow branches, yielding a final collection of 120 programs for our evaluation.

For this task, our objective is to evaluate LLMs on predicting path feasibility and identifying division-by-zero bugs; hence, the execution paths must include both feasible and infeasible cases. To this end, we propose a CFG traversal approach for generating potential execution paths. We first parse each program into a control flow graph (CFG) using the Scalpel static analysis library \cite{li2022scalpel}. The resulting CFG can be represented as a tuple $\mathcal{G} = \{ \mathcal{V}, \mathcal{E}, s, \mathcal{T} \}$, where $\mathcal{V}$ denotes the set of nodes, $\mathcal{E}$ the set of edges, $s$ the entry node corresponding to the start of execution, and $\mathcal{T}$ the set of terminal nodes marking the end of execution. In a Scalpel-generated CFG, each node $v_{i} \in \mathcal{V}$ corresponds to a basic code block, with its final statement potentially being a branch construct (e.g., \texttt{if}, \texttt{for}, \texttt{while}). 

For each program, we perform a depth-first traversal (DFS) on its CFG, beginning from the start node $s$. During the traversal, we record the branch conditions taken at the end of each code block as well as the number of loop iterations. The traversal terminates once all basic blocks have been covered by the collected paths, or when a predefined maximum number of paths is reached. The detailed path traversal algorithm is demonstrated in Algorithm~\ref{alg:traverse}. We set the maximum number of generated paths to 50, the max path length to 100 basic blocks, and the max number of loop iterations to 2.

\begin{algorithm}
        \caption{CFG traversal for extracting execution paths.}\label{alg:traverse}
        
        \begin{algorithmic}[1] % [1] adds line numbers
            \Input {CFG $\mathcal{G} = \left\{ \mathcal{V}, \mathcal{E}, s, \mathcal{T} \right\}$, max path length $L$, max number of paths $N$, max number of loop iteration $K$.}
            \Output {Collected execution paths $\mathcal{P}$}

            \State $\mathcal{P} = \varnothing $, visited nodes $visited = \varnothing $, current node $cur\_node = s$, current path $cur\_path = []$
            
            \Function{visit}{$cur\_node$, $visited$, $cur\_path$}   % Function name and parameter
              \While{$\left | \mathcal{P} \right | < N$ \textbf{and} $\left | visited \right | < \left | \mathcal{V} \right |$}
                \If{$cur\_node \in \mathcal{T}$ \textbf{or} $|cur\_path| > L$}
                  \State $\mathcal{P}$.add($cur\_path$)
                  \Return
                \Else 
                  \State $cur\_path$.append($cur\_node$)
                  \State $visited$.add($cur\_node$)
                \EndIf
                \If{$cur\_node$ match \texttt{for i in range(x):}}
                  \If{\Call{eval\_loop}{value of \texttt{i}, x} == TRUE}
                    \State \Call{visit}{$cur\_node$->true, $visited$, $cur\_path$}
                  \ElsIf{\Call{eval\_loop}{value of \texttt{i}, x} == FALSE}
                    \State \Call{visit}{$cur\_node$->false, $visited$, $cur\_path$}
                  \Else
                    \State \Call{visit}{$cur\_node$->true, $visited$, $cur\_path$}
                    \State \Call{visit}{$cur\_node$->false, $visited$, $cur\_path$}
                  \EndIf
                \Else
                  \For {$node$ in $cur\_node$.successors}
                    \State \Call{visit}{$node$, $visited$, $cur\_path$}
                  \EndFor
                \EndIf
              \EndWhile
            \EndFunction
            
            \State \Call{visit}{$cur\_node$, $visited$, $cur\_path$}
        \end{algorithmic}

    \end{algorithm}

After we collect all the candidates paths, we manually annotate the label for each path based on their feasibility and bug-triggering ability. Each path is categorized into one of the following 3 types of labels:

\begin{itemize}
    \item Valid: This means a path can be satisfied without conflicts on path constraints, and the path does not trigger any ``divided by 0'' bugs.

    \item Invalid: This means a path is infeasible, i.e., cannot find a group of valid inputs of the program to satisfy the given path constraints.

    \item ZeroDivision: This means a path is valid \textbf{until} it triggers a possible ``divided by 0'' bug. 
\end{itemize}

\added{Two authors of this paper annotated the dataset independently and discussed to resolve disagreements. The Cohen's Kappa between the two annotators is 0.82}. After annotating all paths, we manually inspect each program and its extracted execution paths. If a program is capable of producing an execution path of class $k$ but the path extraction algorithm fails to generate it, we manually construct such a path from the CFG and append it to the algorithm-generated path set. \added{For manually-added paths with ``Valid'' or ``ZeroDivision'' types, they are constructed by first assuming an input value that executes without error / triggers a zero division error, and manually collect its execution path along the CFG. For manually-added paths with ``Invalid'' type, they are created by performing random traversal on the CFG, and selecting the paths with constraints unsatisfiable with any inputs before running into any zero division errors.} In the end, we obtain a dataset containing 2,010 execution paths, \added{and the final distribution is shown in Table \ref{tab:classification-labels}. Among all paths, only 79 of them are manually generated, which means these paths do not introduce significant bias towards the distribution of path types.} \deleted{of which 307 are valid, 258 are invalid, and 1,445 exhibit ``division by zero'' bugs.}

\begin{table}[htbp]
\captionof{table}{\added{Label distribution of the path classification dataset.}}
    \label{tab:classification-labels}
    \begin{tabular}{cccc}
      \toprule
      \added{Label} & \added{Count (extracted)} & \added{Count (manual)} & \added{Count (total)}\\
      \midrule
      \added{Valid} & \added{245} & \added{62} & \added{307}\\
      \added{Invalid} & \added{252} & \added{6} & \added{258}\\
      \added{ZeroDivision} & \added{1434} & \added{11} & \added{1445}\\
      \bottomrule
  \end{tabular}
    
\end{table}

\subsubsection{Evaluation Pipeline}

We treat the path classification task as a 3-class classification task. In our setting, we provide the extracted execution path to the LLM, and ask it to generate exactly one class label in ``Valid'', ``Invalid'' and ``ZeroDivision''. The complete prompt template is shown in Figure~\ref{fig:template} (b). For evaluation, we report overall accuracy, \added{balanced accuracy} as well as precision, recall, and F1 scores across all 3 classes \added{and macro setting}. In addition, we measure the number of buggy programs detected: specifically, if the LLM classifies at least one path as ``ZeroDivision,'' we identify the program as buggy.

\iffalse
\begin{figure}[htbp]
\centering
\begin{tcolorbox}[listing only, listing options={basicstyle=\ttfamily}, colback=white, colframe=black,
                  width=\linewidth, sharp corners, boxrule=0.8pt,
                  fonttitle=\bfseries, fontupper=\ttfamily\small]

Here is the execution path for a Python code snippet. You need to determine whether the execution path is valid (i.e., all branch conditions can be satisfied) and whether the path is possible to trigger a 'divided by zero' bug. The execution path consists of a series of code blocks and branch conditions. You should pay extra attention to the last statement for each code block, because it is likely to be a branch condition statement.
You should only return one of the following answers:

- `VALID', which means the execution path is valid and no `divided by zero' bug is triggered.

- `INVALID', which means the execution path is invalid.

- `ZERODIVISION', which means the execution path is valid UNTIL a `divided by zero' bug is triggered.

Only return the answer keyword without any other contents.

<Execution path>

\{\{ Execution path extracted from CFG \}\}

</Execution path>

\end{tcolorbox}
\caption{An Example of the prompt for the path classification task for bug detection.}
\label{fig:template-bug}
\end{figure}
\fi

\subsection{Test Case Generation (real-world)}

\subsubsection{Benchmark Construction}

\added{Test case generation on real-world repositories poses significant challenges to symbolic execution tools for Python, such as PyExSMT \cite{PyExSMT} and CrossHair \cite{CrossHair}. For example, CrossHair requires that all input arguments of a function must be annotated with explicit type information, so it cannot be directly run on real-world repositories where type annotations are often missing. Moreover, for data types that CrossHair cannot handle, especially developer-defined data types, it will directly treat them as \texttt{int} or \texttt{str} variables. As for PyExSMT, it cannot handle external function calls, classes, or exceptions, and does not fully support complex data structures such as lists and dictionaries \cite{xu2025identifying}. These limitations make PyExSMT not suitable for real-world repositories, where classes, external API usage, and exception handling are common.}

For the real-world test case generation task, we aim to evaluate the ability of LLMs in achieving high test coverage by generating test cases that cover specific execution paths. We employ the TestGenEval dataset \cite{jain2025testgeneval}, a large-scale benchmark for test case generation, constructed from 11 public Python repositories. Each data sample in TestGenEval is a code–test file pair, where the code file is the file under test, and the corresponding test file contains test cases that (partially) cover the code file. In its original setting, TestGenEval evaluates LLMs by prompting them to either generate new test files or complete existing ones given the code file under test.

The original TestGenEval focuses on file-level testing, while fine-grained testing on functions, which is important for symbolic execution, has been overlooked. To construct the dataset for our study, we first executed the provided gold test cases in TestGenEvallite (a smaller subset of TestGenEval) and identified functions that were only partially covered by the gold tests. \added{Then, we remove private functions, and functions that cannot be successfully executed due to Git or Docker errors in the testing environment.} From the 160 files under test in TestGenEvallite, we collected \textbf{82 files} with \textbf{\deleted{605}\added{255} partially covered functions}. For each such function, we apply Algorithm~\ref{alg:traverse} to extract up to 10 execution paths (the max path length and max loop iterations are set to the same as the path classification task).

\subsubsection{Evaluation Pipeline}

For each function under test, we prompt the LLM to generate a test case based on all execution paths extracted from its CFG. We adopt the test case completion setting of TestGenEval, where the entire code file containing the function under test, along with the corresponding test file, is provided as context. The complete prompt template is presented in Figure~\ref{fig:template} (c).

After obtaining the LLM-generated test cases for all functions under test, we measure the Pass@1 (whether the generated test can be successfully executed) and line coverage of each function using the \texttt{pytest} \cite{pytest} reports produced by the TestGenEval pipeline. We then compare the coverage achieved by the generated tests against the original gold test files, and a straightforward baseline in which the LLM is prompted to generate tests for the function without being provided with execution paths (similar to the original TestGenEval prompt template, except that we explicitly instruct the LLM to target a specific function).

\iffalse
\begin{figure}[htbp]
\centering
\begin{tcolorbox}[listing only, listing options={basicstyle=\ttfamily}, colback=white, colframe=black,
                  width=\linewidth, sharp corners, boxrule=0.8pt,
                  fonttitle=\bfseries, fontupper=\ttfamily\small]

Below is a code file:

```python

\{code file\}

```

And the current unit test file

```python

\{test file\}

```
\

Your job is to generate a unit test case for function `\{function name\}`, the generated test case will be written to the test file. Your generated test case should cover the execution paths for the function, which are given below:

[Begin of execution paths]

{all execution paths}

[End of execution paths]

You should analyze all given execution paths, and generate test inputs that cover all the satisfiable paths. If a given path is not satisfiable, you should ignore it.
Ideally your generated test case should improve the coverage of the existing test file for the code file.

Only output the generated unit test, preserve indentation and formatting. Do not output anything else. Format like this:

```python

Next unit test Python code

```
\end{tcolorbox}
\caption{The prompt template for the real-world tet case generation task.}
\label{fig:template-testgeneval}
\end{figure}
\fi

\subsection{Experiment Settings}

\subsubsection{Selection of LLMs}
We select 14 popularly adopted LLMs for general complex reasoning and code generation tasks. These LLMs cover non-reasoning and reasoning models:

\begin{itemize}
    \item Non-reasoning LLMs: For proprietary LLMs, we select the GPT-4.1 series of models \cite{gpt-4.1}: GPT-4.1, GPT-4.1-mini, and GPT-4.1-nano. For open-source LLMs, we select DeepSeek-V3 \cite{liu2024deepseek}, Qwen3 (parameter size 30B and 235B) \cite{yang2025qwen3}, and Gemma3 (parameter size 12B and 27B) \cite{team2025gemma}. We also include the state-of-the-art coding LLM Qwen3-coder (the coding version of Qwen 3).

    \item Reasoning models: for large reasoning models (LRMs), we choose proprietary LRMs o3-mini and o4-mini \cite{o3o4}. We also run open-source LRMs, including DeepSeek-R1 \cite{guo2025deepseek} (the reasoning version of DeepSeek-V3), Qwen3-thinking (the reasoning version of Qwen3), and GPT-oss \cite{gpt-oss}.
\end{itemize}

\subsubsection{Experiment configurations}
For open-source LLMs, the DeepSeek models are called using their official API, while we run other models via the OpenRouter \cite{openrouter} platform. For all models that support temperature settings, their temperature is set to 0. The OpenAI LRMs o3-mini and o4-mini support setting different reasoning efforts, and we set them to ``medium''.

%% file: sections/evaluation.tex
\subsection{RQ1: Results on Test Case Generation}

\begin{table}[h]
\centering
\caption{Results (\%) on test case generation from execution paths on competition-level programs.}
  \label{tab:path2test}
  \scalebox{0.8}{
  \begin{tabular}{cllccc}
    \toprule
    Model type & Model & size & Path acc & Node acc (statement) & Node acc (branch)\\
    \midrule
    \multirow{9}{*}{\makecell{Non-reasoning\\ LLM}}& GPT-4.1-nano & N/A & 14.5 & 21.7 & 20.1 \\
    & GPT-4.1-mini & N/A & 37.5 & 37.5 & 37.2 \\
    & GPT-4.1 & N/A & 46.8 & 47.0 & 46.3 \\
    & DeepSeek-v3 & 685B & \added{34.8} & \added{39.5} & \added{39.5} \\
    & \multirow{2}{*}{Gemma3} & 12B & \added{11.4} & \added{18.9} & \added{17.3}\\
    & & 27B & 17.1 & 25.1 & 25.0\\
    & \multirow{2}{*}{Qwen3} & 30B & \added{36.5} & \added{34.9} & \added{34.3}\\
    & & 235B & 45.6 & 41.9 & 42.8 \\
    & Qwen3-coder & 480B & \added{36.2} & \added{38.6} & \added{38.0}\\
    \midrule
    \midrule
    \multirow{5}{*}{LRM} & o3-mini & N/A & \added{52.1} & \added{45.5} & \added{45.0} \\
    & o4-mini & N/A & \textbf{65.6} & 57.8 & 58.2 \\
    & DeepSeek-R1 & 685B & 64.4 & \textbf{60.2} & \textbf{61.5} \\
    & GPT-oss & 120B & \added{54.4} & \added{45.8} & \added{44.1} \\
    & Qwen3-thinking & 235B & 65.0 & 52.6 & 51.9\\
  \bottomrule
\end{tabular}
}
\end{table}

%finding 1: general performance
%different model size, coding vs non-coding
%finding 2: non-reasoning vs reasoning models

Table~\ref{tab:path2test} reports the results of test case generation on competition-level programs. Under our highly challenging setting—where the average execution path exceeds 100 branches—state-of-the-art LLMs still achieve path accuracies above 60\%. The path accuracy varies drastically across different LLMs: small non-reasoning LLMs (e.g., Gemma3-12B and GPT-4.1-nano) have accuracies below 15\%, whereas the best-performing model, o4-mini, achieves 65.6\%. Notably, LRMs consistently outperform non-reasoning LLMs, as none of the non-reasoning models surpass any of the five LRMs in our evaluation. The only coding LLM, Qwen3-coder, does not achieve high performances on all accuracy metrics: its accuracies fall short of Qwen3-235B (\added{its path accuracy is even lower than Qwen3-30b}), despite Qwen3-coder having a much larger parameter size (480B). This finding indicates that while post-training on code corpora enhances LLMs’ capabilities in code generation, it does not necessarily improve their ability to reason about program execution paths.

As for node accuracy, the variance across models is smaller than that observed for path accuracy. For weaker LLMs, their node accuracies are generally higher than path accuracy, indicating that although these models lack the capability to consistently perform correct reasoning over complete execution paths, they are still able to partially capture the constraints of long paths. In contrast, strong LRMs exhibit the opposite trend: their node accuracies are often lower. This suggests that while these models tend to perform well on shorter paths, their errors on extremely long paths with a large number of nodes can negatively impact overall node accuracies.

%ablation study
\begin{table}[htbp]
\renewcommand{\arraystretch}{1.2} % 适当调整行高，让数据不拥挤
\captionof{table}{\added{The ablation study results on removing certain path taggings.}}
\label{tab:ablation_path}
% 使用 resizebox 强制自适应页面宽度
\resizebox{\textwidth}{!}{
% 定义 10 列：前 3 列左对齐，中间 3 列数据，1 个空列 (c)，最后 3 列数据
\begin{tabular}{lll ccc c ccc}
  \toprule
  \multirow{2}{*}{\added{Model type}} & \multirow{2}{*}{\added{Model}} & \multirow{2}{*}{\added{Size}} & \multicolumn{3}
  {c}{\added{w/o loop counter}} & & \multicolumn{3}{c}{\added{w/o loop counter and branch condition}} \\
  \cmidrule(lr){4-6} \cmidrule(lr){8-10}
   & & & \added{P acc} & \added{S acc} & \added{B acc} & & \added{P acc} & \added{S acc} & \added{B acc} \\
  \midrule

  \multirow{9}{*}{\makecell[l]{\added{Non-reasoning} \\ \added{LLM}}}
   & \added{GPT-4.1} & \added{N/A} & \added{37.9 (-8.9)} & \added{36.7 (-10.3)} & \added{35.3 (-11.0)} & & \added{13.9
  (-32.8)} & \added{21.6 (-25.4)} & \added{20.0 (-26.3)} \\
   & \added{GPT-4.1-mini} & \added{N/A} & \added{30.3 (-7.2)} & \added{29.7 (-7.8)} & \added{28.7 (-8.5)} & &
  \added{14.2 (-23.4)} & \added{22.4 (-15.1)} & \added{21.8 (-15.4)} \\
   & \added{GPT-4.1-nano} & \added{N/A} & \added{10.8 (-3.7)} & \added{18.0 (-3.7)} & \added{16.1 (-4.0)} & & \added{6.5
  (-8.0)} & \added{17.6 (-4.1)} & \added{17.1 (-3.0)} \\
   & \added{DeepSeek-v3} & \added{685B} & \added{32.8 (-2.0)} & \added{36.2 (-3.3)} & \added{35.2 (-4.3)} & &
  \added{12.4 (-22.4)} & \added{21.9 (-17.6)} & \added{21.2 (-18.3)} \\
   & \multirow{2}{*}{\added{Gemma3}} & \added{12B} & \added{10.6 (-0.8)} & \added{18.5 (-0.4)} & \added{16.6 (-0.7)} & &
  \added{5.9 (-5.5)} & \added{17.7 (-1.2)} & \added{16.4 (-0.9)} \\
   & & \added{27B} & \added{11.8 (-5.3)} & \added{20.6 (-4.5)} & \added{18.5 (-6.5)} & & \added{7.3 (-9.8)} &
  \added{18.2 (-6.9)} & \added{17.5 (-7.5)} \\
   & \multirow{2}{*}{\added{Qwen3}} & \added{30B} & \added{26.1 (-10.4)} & \added{25.3 (-9.6)} & \added{23.4 (-10.9)} &
  & \added{8.6 (-27.9)} & \added{20.1 (-14.8)} & \added{18.9 (-15.4)} \\
   & & \added{235B} & \added{45.8 (+0.2)} & \added{44.5 (+2.6)} & \added{42.8 (+0.0)} & & \added{15.3 (-30.3)} &
  \added{24.1 (-17.8)} & \added{22.4 (-20.4)} \\
   & \added{Qwen3-coder} & \added{480B} & \added{30.8 (-5.4)} & \added{35.2 (-3.4)} & \added{34.1 (-3.9)} & &
  \added{12.0 (-24.2)} & \added{23.4 (-15.2)} & \added{22.0 (-16.0)} \\
  \midrule

  \multirow{5}{*}{\added{LRM}}
   & \added{o3-mini} & \added{N/A} & \added{36.4 (-15.8)} & \added{34.4 (-11.1)} & \added{33.2 (-11.8)} & & \added{13.0
  (-39.1)} & \added{25.6 (-19.9)} & \added{22.6 (-22.4)} \\
   & \added{o4-mini} & \added{N/A} & \added{50.3 (-15.3)} & \added{47.8 (-10.0)} & \added{46.7 (-11.5)} & & \added{20.0
  (-45.6)} & \added{26.6 (-31.1)} & \added{25.6 (-32.6)} \\
   & \added{DeepSeek-R1} & \added{685B} & \added{53.8 (-10.6)} & \added{45.0 (-15.2)} & \added{43.6 (-17.9)} & &
  \added{13.9 (-50.5)} & \added{22.8 (-37.5)} & \added{20.9 (-40.6)} \\
   & \added{GPT-oss-120b} & \added{120B} & \added{33.8 (-20.6)} & \added{26.2 (-19.6)} & \added{25.2 (-18.9)} & &
  \added{11.2 (-43.2)} & \added{17.7 (-28.1)} & \added{16.3 (-27.8)} \\
   & \added{Qwen3-thinking} & \added{235B} & \added{53.6 (-11.4)} & \added{42.9 (-9.8)} & \added{41.7 (-10.2)} & &
  \added{12.0 (-53.0)} & \added{23.8 (-28.9)} & \added{22.3 (-29.6)} \\
  \bottomrule
  \end{tabular}
}
\end{table}

\added{To further approximate the settings of traditional symbolic execution, we perform an ablation study to remove certain tags from the execution path:}
\begin{enumerate}
    \item \added{Removing the loop iteration counter (e.g., \texttt{iteration: 1}) and not providing an explicit number of loop iterations.}

    \item \added{Removing all branch conditions that have been taken (e.g., \texttt{condition: True}, including loop conditions). In this setting, the LLM has to infer which branch has been taken from the execution path.}
\end{enumerate}

\added{Table \ref{tab:ablation_path} reports the results of the ablation study. Two trends emerge clearly. Here ``P acc'', ``S acc'', and ``B acc'' stand for ``Path acc'', ``Node acc (statement)'', and ``Node acc (branch)''. First, removing only the 
loop iteration counter causes moderate but consistent degradation across all models: non-reasoning LLMs lose 4–11 percentage points in 
path accuracy, while large reasoning models (LRMs) are noticeably more affected, but their accuracies are still consistently higher than non-reasoning models. This indicates that without an explicit loop iteration counter, LLMs may lose track of the loop iteration number, which makes them perform incorrect reasoning on loops. The only exception is Qwen3-235B, which shows similar accuracies, suggesting that some powerful LLMs can partially compensate for missing loop state through predicting loop iteration number themselves, but this does not generalize. }

\added{Second, additionally removing the branch-condition tags causes a far more severe collapse in performance: the path accuracy of all LLMs falls below 25\%. This significant drop is reasonable, since without explicit branch condition tags, LLMs can only infer the taken condition of the branch statement from the content of the next executed statement (if the next statement belongs to the ``True'' branch, then the branch condition is \texttt{True}, and vice versa). Critically, node-level accuracy metrics (S acc and B acc) are less affected than p acc in both conditions, consistent with the fact that path-level correctness requires all branching
decisions to be correct simultaneously. Together, these results demonstrate that both the loop iteration counter and the branch-condition tag encode essential symbolic state: without them, LLMs may lose the information needed to track which path is being described.
}

\begin{figure}[h]
  \centering
  \includegraphics[width=0.8\linewidth]{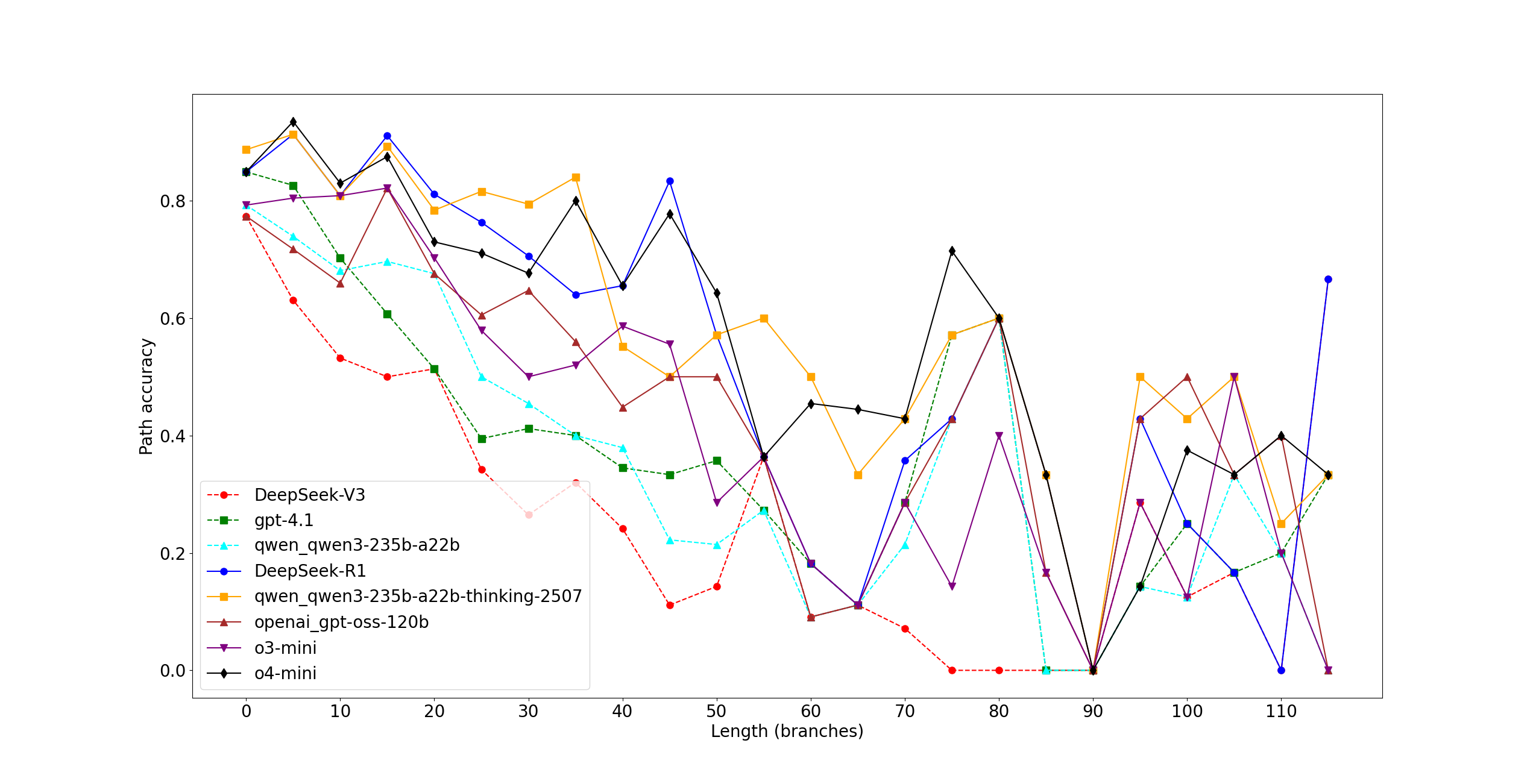}
  \caption{The path accuracy across different execution path lengths (measured by the number of branch conditions).}
  \label{fig:acc_len}
\end{figure}

To better study the relevance between model performance and the execution path length, we demonstrate the path accuracies of various models across varying path lengths in Figure~\ref{fig:acc_len}. We compare the three best-performing non-reasoning models (GPT-4.1, DeepSeek-V3, and Qwen3-235B) against all 5 LRMs in our study. Our results show that for short execution paths with fewer than 10 branch conditions, non-reasoning models achieve accuracies comparable to reasoning models. In medium-length paths (10 $\leqslant$ length $\leqslant$ 50), the advantage of LRMs against non-reasoning LLMs is prominent: state-of-the-art LRMs consistently maintain path accuracies above 0.6, while the performance of non-reasoning models drops rapidly, falling below 0.4 on paths exceeding 30 branches. However, for long paths with more than 50 branch conditions, the accuracies for both reasoning and non-reasoning models become unstable: their path accuracies are lower than 0.4 on most path lengths, and even the strongest models fail to significantly outperform weaker LLMs.

\begin{tcolorbox}[size=title,breakable]
\textbf{Finding 1:} \textcolor{black}{State-of-the-art LLMs are able to generate correct test cases for up to 65\% of execution paths with complex control flows. Overall, LRMs consistently outperform non-reasoning LLMs in test case generation for path constraint solving. In particular, their primary advantage emerges on medium-length execution paths.}
\end{tcolorbox}

\begin{figure}[h]
  \centering
  \includegraphics[width=\linewidth]{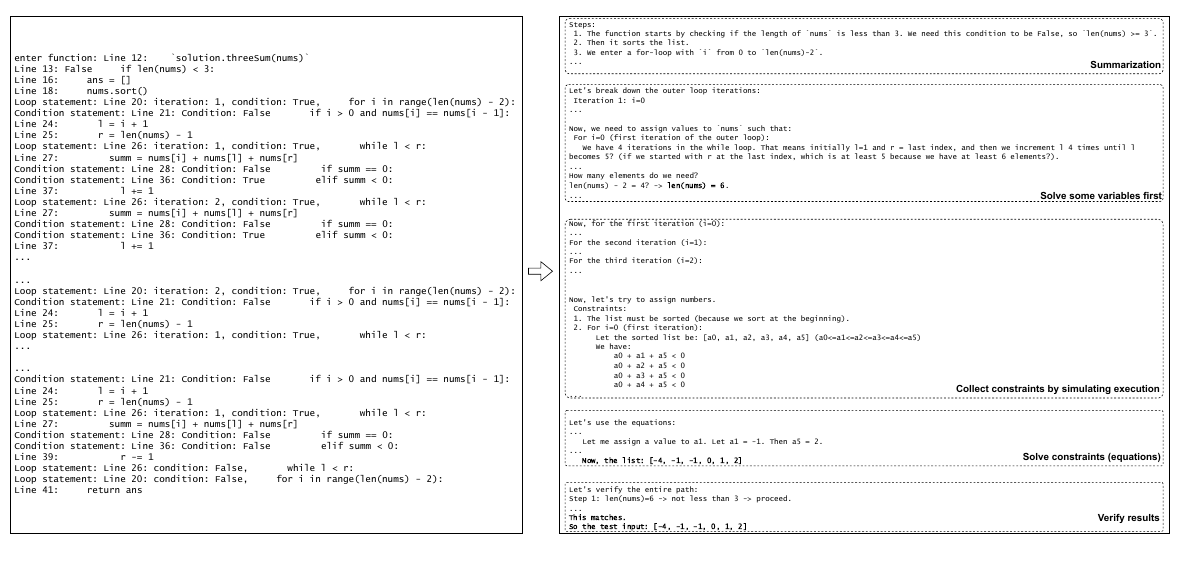}
  \caption{An example of the LRM DeepSeek-R1 successfully generates a correct test case for the given execution path (left). The reasoning CoT is demonstrated on the right.}
  \label{fig:case_testeval}
\end{figure}

\textbf{Study on reasoning CoT}. According to previous studies, the CoT generated by LLMs can provide an approximate view of a model’s reasoning behavior \cite{jie2024interpretable}. Figure~\ref{fig:case_testeval} shows the reasoning CoT generated by DeepSeek-R1 when the LRM generates a correct test case that satisfies the given execution path on the left. In this example, we observe that DeepSeek-R1 follows a five-stage process in generating the test case: (1) the model first summarizes the given execution path; (2) it identifies certain variables that can be determined before others. For instance, in Figure~\ref{fig:case_testeval}, the model infers the length of the input list (6) by counting the number of iterations; (3) it collects all path constraints by simulating the program step by step along the given path; (4) it tries to solve the collected constraints. For complex constraints, the model may adopt trial-and-error strategies, such as assigning tentative values to variables before resolving them; and (5) once all input variables are determined, it validates the correctness of the generated test case by re-simulating execution. If the simulated execution path is not consistent with the given one, the model may retry solving the input variables. This multiple-step reasoning process in LRMs make them more powerful than non-reasoning LLMs in handling complex path constraints.

To further investigate the weaknesses of state-of-the-art LRMs in generating test cases from execution paths, we conduct a human study on the reasoning CoT produced by these models. We focus on DeepSeek-R1, as it is the only LRM that provides full access to its reasoning CoT through an official API. Among the 180 execution paths that DeepSeek-R1 failed to solve, we randomly sample 50 and manually analyze the error patterns in their reasoning CoTs. Errors are classified according to the taxonomy proposed in a prior study \cite{he2025can}. For each execution path, \deleted{we} \added{two authors of this paper} carefully examine its CoT step by step and identify the first error that leads to the incorrect test case. Our analysis reveals the following categories of errors in CoT for test case generation:

\begin{itemize}
    \item \textbf{Reasoning error:} Errors in the logical flow of problem-solving steps.

    \item \textbf{Understanding error:} Incorrect interpretation of programming concepts.

    \item \textbf{Factual error:} Incorrect recall of the input execution path.

    \item \textbf{Calculation error:} Inaccuracies in arithmetic operations or numerical computations.

    \item \textbf{Summary error:} Inadequate conclusion of findings. For example, the reasoning CoT steps are correct, but the generated final answer is wrong.

    \item \textbf{Premise error:} Invalid or incorrect assumptions in the reasoning process.
\end{itemize}

In addition to these basic error types, we also define several sub-types that are related to program language-specific features, such as errors in reasoning/understanding loops, recursion, inline branches (ternary operators `\texttt{if}' and list comprehensions `\texttt{for}'), and external APIs. 
\added{The two annotators reached the Cohen's Kappa of 0.77 on fine-grained sub-types, and a high value of 0.94 on error types, indicating a high level of agreement. For the samples with disagreements, the annotators resolved them through discussion until a consensus was reached.}
The final human evaluation results are shown in \deleted{Figure}\added{Table}~\ref{tab:error-cot}.

\begin{table}[h]
\centering
\caption{Classification results on the error types of DeepSeek-R1 reasoning CoTs in test case generation.}
  \label{tab:error-cot}
  \scalebox{0.7}{
  \begin{tabular}{llcc}
    \toprule
    Error type & Error subtype & Description & Count\\
    \midrule
    \multirow{4}{*}{Reasoning error} & Logic & Generic logic errors, not code-specific & \deleted{5}\added{6}\\
    & Recursion & Error in reasoning on recursive function calls & \deleted{4}\added{3} \\
    & Function call & Error in reasoning on the parameter passing in function calls & \deleted{1}\added{3} \\
    & Class & Error in reasoning on class attributes & 1 \\
    \midrule
    \multirow{4}{*}{Understanding error} & loop & Incorrect interpretation of loop iterations & \deleted{8}\added{7}\\
    & Inline branches & Incorrect interpretation of ternary operators and list comprehensions & \deleted{7}\added{6}\\
    & API & Misunderstanding of Python library APIs & \deleted{12}\added{13}\\
    & Operator & Misunderstanding of Python built-in operators & 1\\
    \midrule
    Factual error & Condition & Error in memorizing branch conditions & \deleted{6}\added{5} \\
    \midrule
    Calculation error & / & / & 1 \\
    \midrule
    Summary error & / & / & 2 \\
    \midrule
    Premise error & / & / & 2 \\
    \bottomrule
\end{tabular}
}
\end{table}

From our human evaluation results, we find that the most common error type in CoT reasoning for test case generation is the misunderstanding of APIs. Most of the APIs used in our evaluation dataset relates to complex data structures, such as queue, min-heap, or other APIs in the Python \texttt{collections} library. This indicates that although large language models can handle external API calls where traditional symbolic execution tools typically fail, their ability to fully understand and reason about such APIs still requires improvement. Other common error types include misunderstanding of loops and \deleted{inline branches} \added{ternary operators}. Reasoning errors are also prevalent, often stemming from incorrect reasoning in logic or complex dependencies, such as recursion and \deleted{non-recursive} function calls.

\begin{tcolorbox}[size=title,breakable]
\textbf{Finding 2:} \textcolor{black}{LRMs are capable of advanced multi-stage reasoning when solving path constraints, which makes them more effective at generating correct test cases than non-reasoning models. Their primary sources of failure lie in difficulties with understanding complex programming constructs (e.g., APIs and loops) and reasoning about program logic.}
\end{tcolorbox}

\subsection{RQ2: Results on Path Classification}

\begin{table}[h]
\centering
\caption{Results on the path classification dataset. The classification accuracy and P/R/F1 scores are shown in \%.}
  \label{tab:bugdetect}
  \resizebox{\textwidth}{!}{
  \begin{tabular}{lccccccccccccccccc}
    \toprule
    \multirow{2}{*}{Model} & \multirow{2}{*}{Size} & \multirow{2}{*}{Acc} & \multirow{2}{*}{\added{Balanced Acc}} & \multicolumn{3}{c}{\added{Macro}} & \multicolumn{3}{c}{Valid} & \multicolumn{3}{c}{Invalid} & \multicolumn{3}{c}{ZeroDivision} & \multirow{2}{*}{No. bugs detected}\\
    \cmidrule{5-7} \cmidrule{8-10} \cmidrule{11-13} \cmidrule{14-16}
    & & & & \added{P} & \added{R} & \added{F1} & P & R & F1 & P & R & F1 & P & R & F1 & \\
    \midrule
    GPT-4.1-nano & N/A & 32.4 & \added{36.8} & \added{39.2} & \added{36.8} & \added{29.3} & 18.8 & 41.0 & 25.8 & 12.4 & 40.7 & 19.0 & 86.9 & 29.1 & 43.5 & 78\\
    GPT-4.1-mini & N/A & 47.8 & \added{51.0} & \added{46.5} & \added{51.0} & \added{41.7} & 30.6 & 73.3 & 43.1 & 15.5 & 34.9 & 21.4 & 93.4 & 44.7 & 60.5 & 93 \\
    GPT-4.1 & N/A & 61.3 & \added{57.4} & \added{50.2} & \added{57.4} & \added{47.4} & 33.4 & \textbf{95.4} & 49.5 & 26.1 & 14.3 & 18.5 & 91.1 & 62.5 & 74.1 & 105\\
    DeepSeek-v3 & 685B & 53.7 & \added{55.4} & \added{49.3} & \added{55.4} & \added{46.2} & 28.5 & 44.6 & 34.8 & 24.5 & 68.6 & \textbf{36.1} & 94.8 & 53.0 & 68.0 & 107\\
    \multirow{2}{*}{Gemma 3} & 12B & 68.3 & \added{60.4} & \added{54.2} & \added{60.4} & \added{55.7} & 47.5 & 73.3 & 57.6 & 23.8 & 34.9 & 28.3 & 91.4 & 73.2 & 81.3 & 104\\
     & 27B & 43.7 & \added{46.7} & \added{43.9} & \added{46.7} & \added{32.4} & 22.3 & 97.1 & 36.3 & 18.0 & 3.5 & 5.9 & 91.5 & 39.5 & 55.2 & 79\\
    \multirow{2}{*}{Qwen3} & 30B & 71.0 & \added{40.5} & \added{43.9} & \added{40.5} & \added{38.4} & 30.3 & 3.3 & 5.9 & 19.4 & 24.8 & 21.8 & 82.1 & 93.6 & 87.5 & 118\\
     & 235B & \textbf{82.9} & \added{62.8} & \added{60.8} & \added{62.8} & \added{58.9} & \textbf{63.0} & 86.0 & \textbf{72.7} & \textbf{28.8} & 6.6 & 10.7 & 90.5 & \textbf{95.9} & \textbf{93.1} & \textbf{120}\\
    Qwen3-coder & 480B & 57.9 & \added{51.1} & \added{50.4} & \added{51.1} & \added{47.5} & 43.1 & 39.4 & 41.2 & 18.6 & 50.8 & 27.2 & 89.6 & 63.0 & 74.0 & 113\\
    \midrule
    o3-mini & N/A & 64.5 & \added{59.3} & \added{52.4} & \added{59.3} & \added{53.2} & 37.5 & 70.0 & 48.8 & 28.0 & 40.3 & 33.0 & 91.8 & 67.6 & 77.9 & 117 \\
    o4-mini & N/A & 43.2 & \added{50.2} & \added{48.1} & \added{50.2} & \added{39.1} & 31.9 & 31.6 & 31.8 & 18.7 & 79.8 & 30.3 & 93.9 & 39.1 & 55.2 & 93\\
    DeepSeek-R1 & 685B & 36.2 & \added{51.1} & \added{50.7} & \added{51.1} & \added{35.1} & 35.6 & 30.3 & 32.8 & 18.4 & 96.5 & 30.9 & 97.5 & 26.6 & 41.9 & 77 \\
    GPT-oss & 120B & 45.1 & \added{51.9} & \added{49.5} & \added{51.9} & \added{40.8} & 34.3 & 32.6 & 33.4 & 19.5 & 81.8 & 31.5 & 94.9 & 41.3 & 57.5 & 104\\
    Qwen3-thinking & 235B & 29.1 & \added{47.6} & \added{52.5} & \added{47.6} & \added{30.4} & 41.0 & 28.3 & 33.5 & 16.8 & \textbf{97.3} & 28.6 & \textbf{99.6} & 17.1 & 29.2 & 64\\
    \midrule
    \added{Base: majority} & \added{N/A} & \added{71.9} & \added{33.3} & \added{24.0} & \added{33.3} & \added{27.9} & \added{0} & \added{0} & \added{0} & \added{0} & \added{0} & \added{0} & \added{71.9} & \added{100.0} & \added{83.7} & \added{120} \\
  \bottomrule
\end{tabular}
}
\end{table}

%finding 4: general performance
Table~\ref{tab:bugdetect} presents the results of the path classification dataset for RQ2. \added{Given that our dataset is highly imbalanced, we introduce a Majority-class Baseline (which trivially predicts 'ZeroDivision' for all paths) and report Macro-F1 and Balanced Accuracy metrics to provide a more rigorous evaluation.} In terms of accuracy, we find that, similar to the test case generation task, state-of-the-art LLMs achieve strong performance on path classification. Qwen3-235B achieves the highest classification accuracy of 82.9\% and an F1 score above 0.93 for detecting paths that trigger division-by-zero bugs. In detecting valid paths, it also achieves the highest F1 of 0.73. Notably, it is the only model in our experiments that successfully identifies all 120 division-by-zero bugs. Most other LLMs achieve accuracies ranging between 45\% and 70\%, suggesting that for many models, our path classification task is less challenging than test case generation. \added{While the Majority-class Baseline achieves a deceptively high overall accuracy of approximately 72\%, its Macro-F1 score is extremely low due to its complete failure to identify valid or invalid paths. In contrast, the evaluated LLMs exhibit genuine learning capabilities by achieving much higher Macro-F1 scores, demonstrating true path-aware reasoning rather than simply exploiting the dataset's class distribution.} Interestingly, we also observe that model size does not necessarily correlate with classification performance. For example, Qwen3-235B outperforms Qwen3-30B in both accuracy and F1 (particularly for detecting valid paths), while Gemma3-27B performs overall worse than Gemma3-12B.
Similar to the test case generation experiment in RQ1, the path classification results of Qwen3-coder are also lower than Qwen3, further proving that the coding ability in LLMs cannot be fully transferred to analyzing execution paths.

Although some LLMs achieve favorable performance in overall path classification accuracy \added{and significantly outperform the trivial baseline in Macro-F1}, their ability to distinguish valid from invalid paths still leaves room for improvement. Except for Qwen3-235B and Gemma3-12B, all other models have an F1 score below 0.5 when detecting valid paths. For invalid paths, the situation is even worse, as no LLM achieves an F1 score above 0.4. These results suggest that standalone LLMs remain insufficiently powerful to reliably differentiate feasible from infeasible paths. In the future, if LLMs are to be used for reducing the execution path search space in symbolic execution by pruning infeasible paths, additional augmentation techniques will be necessary. 

\begin{tcolorbox}[size=title,breakable]
\textbf{Finding 3:} \textcolor{black}{Existing LLMs can achieve up to 83\% accuracy in classification tasks for path constraint solving. This proves the potential of state-of-the-art LLMs in identifying valid execution paths and detecting runtime errors from path constraints. However, the accuracy of most LLMs remains limited in detecting infeasible execution paths.}
\end{tcolorbox}

\begin{figure}[h]
  \centering
  \includegraphics[width=0.9\linewidth]{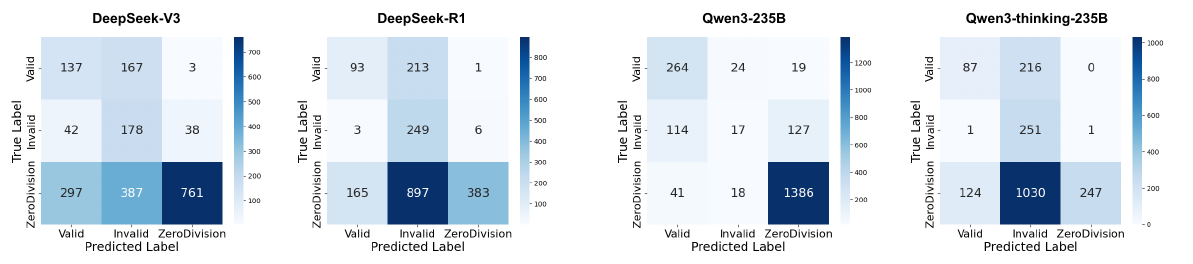}
  \caption{Heatmap of the distribution of the predicted execution path labels vs. the ground-truth labels.}
  \label{fig:bug-lrm}
\end{figure}

However, unlike the test case generation task, where the most powerful LRMs achieve the highest performance, LRMs do not exhibit clear superiority over non-reasoning LLMs in the path classification task. Moreover, we notice that for state-of-the-art LRMs, o4-mini and DeepSeek-R1, their classification accuracy and F1 are surprisingly lower than most LLMs. Most LRMs in our experiment share a similar pattern in their results: they demonstrate higher recall for invalid paths but lower recall on paths with zero division compared to non-reasoning LLMs. Figure~\ref{fig:bug-lrm} illustrates the heatmap of labels predicted by non-reasoning LLMs (DeepSeek-V3 and Qwen3) vs. LRMs (DeepSeek-R1 and Qwen3-thinking). Compared to non-reasoning LLMs, the majority of prediction errors made by LRMs are misclassifying zero-division paths as invalid paths, whereas invalid paths are rarely misclassified as either valid or zero-division.

\begin{figure}[h]
  \centering
  \includegraphics[width=0.8\linewidth]{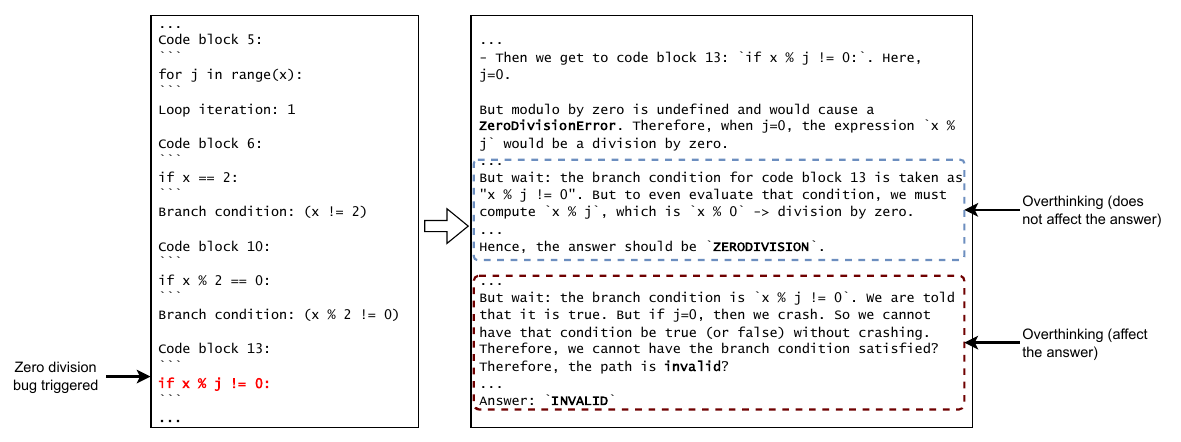}
  \caption{An example of the execution path (left) and reasoning CoT (right, generated by DeepSeek-R1) when the LRM misclassifies a zero division path into an invalid path.}
  \label{fig:case-bug}
\end{figure}

%Figure~\ref{fig:case-bug} shows a case study on the reasoning CoT of a 
Figure~\ref{fig:case-bug} shows a case study on the reasoning CoT of a misclassified execution path. In this execution path, a zero division error is triggered at the statement ``\texttt{x \% j != 0}'' because the variable \texttt{j} should be zero at the first iteration in the loop ``\texttt{for j in range(x)}''. In the reasoning CoT generated by DeepSeek-R1, the model first correctly locate and explain the error. However, the model does not output the final answer at this point. Instead, it continuously generates excessive reasoning steps start with ``But wait''. This phenomenon, referred to as ``overthinking'' \cite{chen2024not}, is a typical behavior widely observed in LRMs. At first, the reasoning steps are still in consistent with the correct answer ``ZERODIVISION''. But then, it gets stuck with the branch condition while the zero-division bug is already triggered, thus incorrectly changes the answer to ``INVALID''. We also notice that when the length of the reasoning CoT increases, the classification accuracy tends to drop: in the 960 paths when the generated CoT is shorter than 3,000 tokens the classification accuracy is 46.8\%, while in the 1,050 paths with DeepSeek-R1-generated CoT longer than 3,000 tokens, the accuracy drops to 26.5\%. \deleted{This highlights that overly long reasoning steps in LRMs not only reduce classification accuracy but also lead to unnecessary token consumption.}

\begin{table}[htbp]
    \caption{\added{The CoT (generated by DeepSeek-R1) length, classification metrics, and correlation analysis between CoT length and accuracy on different path length buckets.}}
  \label{tab:overthinking-path-buckets}
    \scalebox{1}{
  \begin{tabular}{lcccccc}
  \toprule
  \added{Path length range} & \added{Count} & \added{Average CoT length} & \added{Accuracy} & \added{Macro F1} &
  \added{Spearman} & \added{p-value} \\
  \midrule
  \added{[1, 10)} & \added{674} & \added{2224.52} & \added{44.66\%} & \added{0.4270} & \added{-0.4173} & \added{0e-6}
  \\
  \added{[10, 20)} & \added{844} & \added{2968.96} & \added{31.87\%} & \added{0.2795} & \added{-0.1655} &
  \added{1e-6} \\
  \added{[20, MAX)} & \added{492} & \added{2080.70} & \added{31.50\%} & \added{0.2531} & \added{-0.2014} &
  \added{7e-6} \\
  \bottomrule
  \end{tabular}
  }
\end{table}

\begin{center}
\captionof{table}{\added{The CoT (generated by DeepSeek-R1) length, classification metrics, and correlation analysis between CoT length and accuracy on different path difficulty buckets measured by pass rates.}}
  \label{tab:overthinking-path-buckets-passcount}
  \scalebox{1}{
  \begin{tabular}{lcccccc}
  \toprule
  \added{No. LLMs passed} & \added{Count} & \added{Average CoT length} & \added{Accuracy} & \added{Macro F1} &
  \added{Spearman} & \added{p-value} \\
  \midrule
  \added{pass=0-3} & \added{488} & \added{2659.14} & \added{43.65\%} & \added{0.3318} & \added{-0.2175} &
  \added{1e-6} \\
  \added{pass=4-6} & \added{874} & \added{2716.51} & \added{28.60\%} & \added{0.2837} & \added{-0.2357} &
  \added{0e-6} \\
  \added{pass=7-9} & \added{648} & \added{2094.05} & \added{40.43\%} & \added{0.3784} & \added{-0.2895} &
  \added{0e-6} \\
  \bottomrule
  \end{tabular}
  }
  
\end{center}

\begin{center}
\captionof{table}{\added{The classification accuracy and macro F1 in each CoT length interval.}}
  \label{tab:overthinking-cot-buckets}
     \scalebox{1}{
  \begin{tabular}{llcc}
  \toprule
  \added{Path length range} & \added{CoT length range} & \added{Accuracy} & \added{Macro F1} \\
  \midrule
  \multirow{4}{*}{\added{[1,10)}}
  & \added{282-1118}   & \added{84.02\%} & \added{0.6838} \\
  & \added{1127-1701}  & \added{38.10\%} & \added{0.3660} \\
  & \added{1737-2873}  & \added{30.36\%} & \added{0.2963} \\
  & \added{2879-MAX} & \added{26.04\%} & \added{0.2399} \\
  \midrule
  \multirow{4}{*}{\added{[10,20)}}
  & \added{342-1834}   & \added{45.02\%} & \added{0.3884} \\
  & \added{1841-2665}  & \added{31.28\%} & \added{0.2698} \\
  & \added{2667-3785}  & \added{27.01\%} & \added{0.2455} \\
  & \added{3787-MAX} & \added{24.17\%} & \added{0.2033} \\
  \midrule
  \multirow{4}{*}{\added{[20,74)}}
  & \added{257-915}    & \added{49.59\%} & \added{0.3705} \\
  & \added{916-1536}   & \added{24.39\%} & \added{0.1795} \\
  & \added{1542-2771}  & \added{28.46\%} & \added{0.1959} \\
  & \added{2789-MAX}  & \added{23.58\%} & \added{0.2146} \\
  \bottomrule
  \end{tabular}
  }
  
\end{center}

%\todo{Change the text to blue}

\added{
To quantitatively study the correlation between CoT length and classification accuracy when controlling the confounding effect of path difficulty. We group the data by two difficulty measures, namely path length and the pass count of non-reasoning models: for each path, we used the number of LLMs that passed this path (out of 9 in our paper) as a proxy difficulty measure. We split the dataset into 3 buckets by these two difficulty measures, and compute the Spearman's rank correlation coefficient between CoT length and classification accuracy within each bucket. From the results in Table \ref{tab:overthinking-path-buckets} and Table \ref{tab:overthinking-path-buckets-passcount}, we find that longer paths do not necessarily result in longer CoTs. Furthermore, the correlation between CoT length and classification performance remains consistently negative within each bucket, and the $p-value$ is consistently lower than 0.05, suggesting that CoT length is indeed negatively correlated with path classification accuracy, and implies that overthinking may be one of the reasons that LRMs do not significantly outperform non-reasoning models on path classification. These results suggest that the association between longer CoTs and worse classification performance cannot be explained solely by harder paths eliciting longer reasoning. When we take a closer look at each path length bucket in Table \ref{tab:overthinking-cot-buckets}, the finding still stands. In this table, we split the CoTs for each path length range into 4 buckets of the same size by their length. When CoT length increases, the classification accuracy and macro F1 decline.
}

%finding 5: error pattern, preference of predicting valid/invalid/zero across different models?

To further analyze the behavior pattern of LRMs on the path classification task, we ask another question: \textit{Will LRMs perform well when there is no distraction from invalid/buggy paths?} To answer this question, we reformulate the path classification task into two binary classification tasks: classification between valid/invalid paths and classification between valid and zero-division paths.

\begin{table}[h]
\centering
\caption{Results on the path classification dataset with binary classification settings.}
  \label{tab:bugdetect-2class}
  \scalebox{0.7}{
  \begin{tabular}{lccccccccccccccc}
    \toprule
    \multirow{3}{*}{Model}  & \multicolumn{7}{c}{Valid/Invalid} & & \multicolumn{7}{c}{Valid/ZeroDivision}\\
    \cmidrule{2-8} \cmidrule{10-16}
    & \multirow{2}{*}{Acc} & \multicolumn{3}{c}{Valid} & \multicolumn{3}{c}{Invalid} & & \multirow{2}{*}{Acc} & \multicolumn{3}{c}{Valid} & \multicolumn{3}{c}{ZeroDivision}\\
    \cmidrule{3-5} \cmidrule{6-8} \cmidrule{11-13} \cmidrule{14-16}
    & & p & r & f1 & p & r & f1 & & & p & r & f1 & p & r & f1\\
    \midrule
    DeepSeek-v3 & \textbf{60.7} & 89.7 & 31.3 & 46.4 & \textbf{53.9} & 95.7 & \textbf{69.0} & & 79.9 & \textbf{46.5} & 96.1 & \textbf{62.6} & \textbf{99.0} & 76.5 & 86.3\\
    Qwen3-235B & 54.5 & 60.6 & \textbf{46.6} & \textbf{52.7} & 50.2 & 64.0 & 56.2 & & \textbf{80.3} & 45.2 & 58.3 & 50.9 & 90.6 & \textbf{85.0} & \textbf{87.7}\\
    \midrule
    \midrule
    o4-mini & 57.5 & 78.6 & 30.0 & 43.4 & 52.0 & 90.3 & 66.1 & & 66.0 & 34.0 & \textbf{99.7} & 50.7 & 99.9 & 58.8 & 74.0\\
    DeepSeek-R1 & 59.1 & \textbf{90.4} & 27.7 & 42.4 & 52.9 & \textbf{96.5} & 68.3 & & 64.9 & 33.2 & 99.4 & 49.8 & 99.8 & 57.6 & 73.0\\
    GPT-oss & 51.5 & 63.9 & 24.8 & 35.7 & 48.5 & 83.3 & 61.3 & & 62.8 & 26.6 & 62.9 & 37.4 & 88.8 & 62.8 & 73.6\\
    Qwen3-thinking & 51.5 & 70.9 & 19.9 & 31.0 & 48.4 & 89.2 & 62.8 & & 53.9 & 22.8 & 62.2 & 33.4 & 87.7 & 52.1 & 65.4\\
  \bottomrule
\end{tabular}
}
\end{table}

Table~\ref{tab:bugdetect-2class} demonstrates the results of LLMs under the two binary classification settings. For this experiment, we select LRMs that achieved suboptimal performances in the original multi-class setting (namely, DeepSeek-R1, Qwen3-thinking, o4-mini, and GPT-oss), and compare them with their non-reasoning counterparts (only DeepSeek-V3 and Qwen3-235B are available). We find that even if we simplify the path classification task to binary settings, these LRM still fall behind non-reasoning models. Specifically, none of the four LRMs surpasses the two non-reasoning LLMs in either accuracy or F1 when distinguishing valid paths from invalid ones. Likewise, all four LRMs underperform the non-reasoning models in classifying valid versus bug-inducing paths. The LRMs generally tend to misclassify valid paths as invalid in the valid/invalid classification, as indicated by their high recall but low precision on invalid paths. For the valid/buggy classification task, LRMs tend to misclassify zero-division execution paths as valid.

\begin{tcolorbox}[size=title,breakable]
\textbf{Finding 4:} \textcolor{black}{The extensive reasoning process of LRMs may undermine their performance on execution path classification. Even when the task is reduced to binary classification, some LRMs are still outperformed by their non-reasoning counterparts.}
\end{tcolorbox}

\subsection{RQ3: Results on Real-World Repositories}

\begin{table}[h]
\centering
\caption{Results on test case generation on real-world repositories. Pass and coverage metrics are reported in percentages. ``Path'' in column ``Prompt'' means execution paths are provided, ``Baseline'' means no execution paths. The results in brackets are the improvements over the gold test case provided in the dataset \deleted{(for the whole dataset with 605 functions under test, the coverage for gold test cases is 78.4\%)}(\added{for the whole dataset with 255 functions under test, the coverage for gold test cases is 81.1\%).}}
  \label{tab:testgeneval}
  \scalebox{0.7}{
  \begin{tabular}{lcccccccc}
  \toprule
  Model & size & Prompt & Pass@1 & Line cov & Improvement & Line cov@Pass (gold) & Line cov@Pass & Improvement@Pass\\
  \midrule
  \multirow{2}{*}{GPT-4.1-nano} & \multirow{2}{*}{N/A} & Path & \added{9.7} & \added{81.8}(+\added{0.7}) & \multirow{2}{*}{+\added{0.2}} & \added{81.3} &
  \added{94.4}(+\added{13.1}) & \multirow{2}{*}{+\added{8.0}}\\
  & & Base & \added{9.8} & \added{81.6}(+\added{0.5}) & & \added{84.1} & \added{89.2}(+\added{5.1}) \\
  \multirow{2}{*}{GPT-4.1-mini} & \multirow{2}{*}{N/A} & Path & \added{24.3} & \added{83.3}(+\added{2.2}) & \multirow{2}{*}{+\added{0.2}} & \added{84.9}
  & \textbf{\added{97.1}}(+\added{12.2}) & \multirow{2}{*}{+\added{0.8}}\\
  & & Base & \added{23.9} & \added{83.1}(+\added{2.0}) & & \added{81.1} & \added{92.5}(+\added{11.4}) \\
  \multirow{2}{*}{GPT-4.1} & \multirow{2}{*}{N/A} & Path & \added{36.9} & \added{85.9}(+\added{4.8}) & \multirow{2}{*}{+\added{1.1}} & \added{77.0} &
  \added{95.0}(+\added{18.0}) & \multirow{2}{*}{+\added{5.1}}\\
  & & Base & \added{33.3} & \added{84.8}(+\added{3.7}) & & \added{81.3} & \added{94.2}(+\added{12.9}) \\
  \multirow{2}{*}{DeepSeek-v3} & \multirow{2}{*}{685B} & Path & \added{39.6} & \added{83.4}(+\added{2.3}) & \multirow{2}{*}{+\added{1.1}} &
  \added{80.0} & \added{88.1}(+\added{8.1}) & \multirow{2}{*}{+\added{4.6}}\\
  & & Base & \added{36.5} & \added{82.3}(+\added{1.2}) & & \added{79.2} & \added{82.7}(+\added{3.5}) \\
  \multirow{2}{*}{Gemma 3} & \multirow{2}{*}{12B} & Path & \added{5.9} & \added{81.5}(+\added{0.4}) & \multirow{2}{*}{+\added{0.4}} &
  \added{74.4} & \added{86.3}(+\added{11.9}) & \multirow{2}{*}{+\added{10.8}}\\
  & & Base & \added{10.6} & \added{81.1}(+\added{0.0}) & & \added{77.3} & \added{78.4}(+\added{1.1}) \\
  \multirow{2}{*}{Gemma 3} & \multirow{2}{*}{27B} & Path & \added{19.6} & \added{81.8}(+\added{0.7}) & \multirow{2}{*}{+\added{0.3}} &
  \added{82.0} & \added{86.0}(+\added{4.0}) & \multirow{2}{*}{+\added{2.5}}\\
  & & Base & \added{32.9} & \added{81.5}(+\added{0.4}) & & \added{78.7} & \added{80.2}(+\added{1.5})\\
  \multirow{2}{*}{Qwen3} & \multirow{2}{*}{30B} & Path & \added{22.1} & \added{81.6(+0.5)} & \multirow{2}{*}{+\added{0.3}} & \added{67.4} & \added{67.9(+0.5)} & \multirow{2}{*}{+\added{0.3}} \\
  & & Base & \added{15.5} & \added{81.3(+0.2)} & & \added{68.2} & \added{68.4(+0.2)} &  \\
  \multirow{2}{*}{Qwen3} & \multirow{2}{*}{235B} & Path & \added{24.7} & \added{83.0}(+\added{1.9}) & \multirow{2}{*}{+\added{0.2}} &
  \added{83.9} & \added{92.0}(+\added{8.1}) & \multirow{2}{*}{+\added{2.3}}\\
  & & Base & \added{30.2} & \added{82.8}(+\added{1.7}) & & \added{84.5} & \added{90.3}(+\added{5.8})\\
  \multirow{2}{*}{Qwen3-coder} & \multirow{2}{*}{480B} & Path & \added{29.4} & \added{84.2}(+\added{3.1}) & \multirow{2}{*}{+\added{0.8}} &
  \added{81.0} & \added{96.0}(+\added{15.0}) & \multirow{2}{*}{+\added{8.1}}\\
  & & Base & \added{34.5} & \added{83.4}(+\added{2.3}) & & \added{83.5} & \added{90.4}(+\added{6.9})\\
  \midrule
  \midrule
  \multirow{2}{*}{o3-mini} & \multirow{2}{*}{N/A} & Path & \added{38.0} & \added{85.4}(+\added{4.3}) & \multirow{2}{*}{+\added{1.1}} & \added{83.0} &
  \added{96.2}(+\added{13.2}) & \multirow{2}{*}{+\added{4.1}} \\
  & & Base & \added{40.8} & \added{84.3}(+\added{3.2}) & & \added{84.0} & \added{93.1}(+\added{9.1}) \\
  \multirow{2}{*}{o4-mini} & \multirow{2}{*}{N/A} & Path & \textbf{\added{53.7}} & \textbf{\added{86.5}(+\added{5.4})} & \multirow{2}{*}{+\added{0.6}} &
  \added{83.8} & \added{94.7}(+\added{10.9}) & \multirow{2}{*}{+\added{0.5}}\\
  & & Base & \added{51.0} & \added{85.9}(+\added{4.8}) &  & \added{84.4} & \added{94.8}(+\added{10.4}) \\
  \multirow{2}{*}{DeepSeek-R1} & \multirow{2}{*}{685B} & Path & \added{37.3} & \added{84.2}(+\added{3.1}) & \multirow{2}{*}{+\added{1.0}} &
  \added{83.1} & \added{92.4}(+\added{9.3}) & \multirow{2}{*}{+\added{3.6}}\\
  & & Base & \added{39.6} & \added{83.2}(+\added{2.1}) & & \added{83.9} & \added{89.6}(+\added{5.7}) \\
  \multirow{2}{*}{GPT-oss} & \multirow{2}{*}{120B} & Path & \added{27.1} & \added{84.0}(+\added{2.9}) & \multirow{2}{*}{+\added{1.2}} &
  \added{80.7} & \added{95.1}(+\added{14.4}) & \multirow{2}{*}{+\added{3.0}} \\
  & & Base & \added{21.6} & \added{82.8}(+\added{1.7}) & & \added{83.8} & \added{95.2}(+\added{11.4})\\
  \multirow{2}{*}{Qwen3-thinking} & \multirow{2}{*}{235B} & Path & \added{26.3} & \added{83.4}(+\added{2.3}) & \multirow{2}{*}{+\added{0.2}} &
  \added{84.1} & \added{94.3}(+\added{10.2}) & \multirow{2}{*}{+\added{1.5}}\\
  & & Base & \added{29.8} & \added{83.2}(+\added{2.1}) & & \added{82.8} & \added{91.5}(+\added{8.7})\\
  \bottomrule
  \end{tabular}
  }
\end{table}

%finding 6: general coverage performance

Table~\ref{tab:testgeneval} reports the Pass@1 and coverage results for real-world test case generation. We measure the coverage metrics from two perspectives: (i) to what extent LLM-generated test cases improve coverage compared to the gold tests, and (ii) how providing execution path information influences the coverage of generated test cases (with improvements reported in the ``Improvement'' columns). We measure both the overall coverage and the coverage on test cases that can pass the execution (cov@Pass).
\deleted{For all LLMs, providing execution paths consistently improves line coverage for functions that are not fully covered by the gold tests. In contrast, due to the complexity of repository structure and external API calls, traditional symbolic execution tools such as PyExSMT \cite{PyExSMT} and CrossHair \cite{CrossHair} cannot run successfully on this dataset. The best-performing model is o4-mini: it achieved the highest 83.2\% overall line coverage. These results suggest that LLM-based path constraint solving can indeed help real-world software testing tasks.}
\added{For all LLMs with valid path/baseline pairs in Table~\ref{tab:testgeneval}, providing execution paths improves overall line coverage, although the gains are sometimes modest. In contrast, due to the complexity of repository structure and external API calls, traditional symbolic execution tools such as PyExSMT \cite{PyExSMT} and CrossHair \cite{CrossHair} cannot run successfully on this dataset. The best-performing model is o4-mini: it achieved the highest 86.5\% overall line coverage and the highest Pass@1 of 53.7\%. These results further suggest that LLM-based path constraint solving can indeed help software testing in real-world repositories.}

\deleted{Overall, we find that the primary limitation to improving test coverage is the pass rate of generated test cases. In our experiment, the Pass@1 for all models is not higher than 42\%, and execution path information does not improve pass rates. For instance, although GPT-4.1-mini achieves a near-perfect cov@Pass of 97.1\%, its low Pass@1 prevents this strength from transforming into substantial overall coverage gains. When we examine cov@Pass, LRMs and state-of-the-art non-reasoning LLMs (e.g., GPT-4.1 and DeepSeek-V3) can achieve cov@Pass around 95\% after providing execution paths, while for smaller LLMs (e.g., Gemma3-12b), their cov@Pass falls below 90\%. This indicates that state-of-the-art LLMs, no matter reasoning or not, both have strong reasoning abilities in real-world software execution paths. This is different from our observation on the TestEval dataset in RQ1, where LRMs show a significant advantage over non-reasoning models.}
\added{Overall, we find that the primary limitation to improving test coverage is still the executability of generated test cases. The best Pass@1 reaches 53.7\%, but several models still remain below 30\%. Moreover, execution path information improves coverage much more consistently than it improves Pass@1. For instance, GPT-4.1-mini achieves a near-perfect cov@Pass of 97.1\%, but its Pass@1 is only 24.3\%, which limits how much this strength can be translated into overall coverage gains. When we examine cov@Pass, both LRMs and state-of-the-art non-reasoning LLMs can achieve around 95\% after providing execution paths, including GPT-4.1, o3-mini, GPT-oss, and Qwen3-coder, while for smaller LLMs such as Gemma3-12b, cov@Pass remains much lower at 86.3\%. This again indicates that state-of-the-art LLMs, whether reasoning or not, both show some ability to reason about execution paths in real-world software. This is different from our observation on the TestEval dataset in RQ1, where LRMs show a significant advantage over non-reasoning models.}

%mutually covered by path and baseline

\begin{table}[htbp]
\caption{\added{Coverage on the subset of functions both covered by giving execution path and baseline prompts.}}
  \label{tab:testgeneval-mutual}
  \scalebox{1}{
   \begin{tabular}{lcccc}
  \toprule
  \added{Model} & \added{Gold cov} & \added{Cov w/ path} & \added{Cov w/o path} & \added{Improvement w/ path} \\
  \midrule
  \added{GPT-4.1-nano} & \added{79.1\%} & \added{95.7\%(+16.5)} & \added{91.3\%(+12.2)} & \added{+4.4} \\
  \added{GPT-4.1-mini} & \added{82.6\%} & \added{97.7\%(+15.1)} & \added{95.0\%(+12.5)} & \added{+2.7} \\
  \added{GPT-4.1} & \added{72.0\%} & \added{94.3\%(+22.2)} & \added{94.3\%(+22.2)} & \added{+0.0} \\
  \added{DeepSeek-v3} & \added{76.2\%} & \added{82.1\%(+5.9)} & \added{81.5\%(+5.3)} & \added{+0.5} \\
  \added{Gemma 3-12B} & \added{56.3\%} & \added{81.3\%(+25.0)} & \added{68.8\%(+12.5)} & \added{+12.5} \\
  \added{Gemma 3-27B} & \added{76.0\%} & \added{78.8\%(+2.8)} & \added{77.2\%(+1.2)} & \added{+1.6} \\
  \added{Qwen3-30B} & \added{66.8\%} & \added{67.3\%(+0.5)} & \added{67.0\%(+0.2)} & \added{+0.3}\\
  \added{Qwen3-235B} & \added{83.9\%} & \added{91.4\%(+7.5)} & \added{90.1\%(+6.2)} & \added{+1.3} \\
  \added{Qwen3-coder} & \added{82.7\%} & \added{96.6\%(+13.8)} & \added{92.1\%(+9.4)} & \added{+4.5} \\
  \midrule
  \added{o3-mini} & \added{81.3\%} & \added{97.8\%(+16.5)} & \added{95.6\%(+14.3)} & \added{+2.2} \\
  \added{o4-mini} & \added{82.2\%} & \added{97.1\%(+14.9)} & \added{96.2\%(+14.0)} & \added{+0.9} \\
  \added{DeepSeek-R1} & \added{82.8\%} & \added{93.4\%(+10.6)} & \added{91.3\%(+8.5)} & \added{+2.1} \\
  \added{GPT-oss} & \added{74.0\%} & \added{97.6\%(+23.7)} & \added{96.2\%(+22.2)} & \added{+1.5} \\
  \added{Qwen3-thinking} & \added{84.5\%} & \added{95.9\%(+11.4)} & \added{94.5\%(+10.0)} & \added{+1.4} \\
  \bottomrule
  \end{tabular}
  }
  
\end{table}

\added{Table~\ref{tab:testgeneval-mutual} further measures the coverage on functions that are mutually passed for both w/ and w/o execution paths. Providing execution paths consistently improves the coverage over the gold tests, and also yields higher coverage than the baseline-setting prompt for all models except GPT-4.1. The gains over the baseline prompt are marginal for most strong models (because their baselines are already near the coverage upper bound), but can still be substantial for some smaller models, which is consistent with our claim that execution paths mainly help improve coverage quality once tests are executable, while the main bottleneck in real-world repositories remains test executability rather than path-aware reasoning itself.}

\begin{tcolorbox}[size=title,breakable]
\textbf{Finding 5:} \deleted{Execution path can help most LLMs in further improving test coverage in real-world software
  repositories. On this task, state-of-the-art non-reasoning LLMs exhibit reasoning capabilities comparable to those of LRMs. Nevertheless, the primary
  challenge for LLM-based real-world software testing remains the generation of executable test cases.}\added{Execution path can help most LLMs in
  further improving test coverage in real-world software repositories, but does not help LLMs in improving test pass rates. On this task, state-of-the-art non-reasoning LLMs exhibit test coverage capabilities comparable to those of LRMs. }
\end{tcolorbox}

%% file: sections/discussion.tex
%finding 3: time efficiency
\subsection{Analysis on Efficiency} 
When using LLMs to replace traditional program analysis, the time and cost efficiency should also be carefully considered. Table~\ref{tab:efficiency} shows the average running time and tokens consumed (directly relevant to the cost) for LLMs in test case generation for RQ1. As the tokenizer for many LLMs cannot be publicly accessed, we estimate token counts using the OpenAI \texttt{tiktoken} \footnote{https://github.com/openai/tiktoken} library with the tokenizer of GPT-4o. We find that for non-reasoning LLMs and proprietary LRMs with reasoning control (the reasoning effort of o3-mini and o4-mini can be controlled via an API hyperparameter), their running time per path varies from 15 to 60 seconds. Given the lengths of the execution paths, the efficiency of these models is generally favourable. 
However, state-of-the-art open-source LRMs are extremely inefficient on this task, despite achieving higher performance than non-reasoning models. For example, DeepSeek-R1 takes more than 5 minutes to solve a single path, which is unacceptable for large-scale program analysis workloads. The complexity of LRMs is further illustrated by their token consumption: DeepSeek-R1 generates over 7,000 tokens for its reasoning CoT, which is more than 15 times longer than the final output. Looking forward, if open-source LRMs are to be applied to path constraint solving, effective strategies for limiting the reasoning CoT length will be urgently needed.

We also compare the efficiency against traditional symbolic execution tools. We choose the Python symbolic execution tool CrossHair \cite{CrossHair}, and run it on all 210 programs for test case generation. As there is no possible way to use CrossHair to solve a single predefined execution path, we simply run \texttt{crosshair cover} command to generate test cases for all programs with a timeout setting of 600 seconds, and measure the running time of testing each program. On average, CrossHair takes 13.0 seconds to generate test cases for each program (reaching 100\% line coverage). The running time is shorter than 13 out of 14 LLMs in our study for solving a single path. This suggests that the time efficiency of most LLMs still cannot compete with traditional solver-based symbolic execution tools. To make LLMs more practical for path constraint solving, integrating them with traditional tools may yield better efficiency than relying on LLMs alone. However, there is one advantage in the time efficiency of LLMs: it will complete constraint solving in a limited time, no matter the complexity of the execution path, while traditional tools may fail on these complex paths. For example, the longest time for o4-mini to solve a path in RQ1 is 361 seconds, while CrossHair encountered a timeout on this path.

\begin{tcolorbox}[size=title,breakable]
\textbf{Finding 6:} \textcolor{black}{Non-reasoning LLMs demonstrate favorable time and token efficiency in path constraint solving. In contrast, the time efficiency of LRMs without reasoning control is low. Overall, the time efficiency of most LLMs remains lower than that of traditional symbolic execution tools, but LLMs can be more efficient than tools on some extremely complex paths.}
\end{tcolorbox}

\begin{table}[h]
\centering
\caption{The average time (seconds) and tokens consumed for generating a test case for an execution path in RQ1. The results in brackets indicate the length of the generated reasoning CoTs for LRMs.}
  \label{tab:efficiency}
  \scalebox{0.5}{
  \begin{tabular}{lccccccccccccccc}
  \toprule
  \multirow{2}{*}{Model} & \multirow{2}{*}{GPT-4.1-nano} & \multirow{2}{*}{GPT-4.1-mini} & \multirow{2}{*}{GPT-4.1} & \multirow{2}{*}{DeepSeek-v3} & \multicolumn{2}{c}{Gemma3} & \multicolumn{2}{c}{Qwen3} & \multirow{2}{*}{Qwen3-coder} & \multirow{2}{*}{o3-mini} & \multirow{2}{*}{o4-mini} & \multirow{2}{*}{DeepSeek-R1} & \multirow{2}{*}{GPT-oss} & \multirow{2}{*}{Qwen3 thinking} & \multirow{2}{*}{\makecell{CrossHair\\(Program level)}} \\
  & & & & & 12B & 27B & 30B & 235B & & & & & & &\\
  \midrule
  Time (s) & 16.0 & 29.0 & 28.4 & 30.5 & 25.1 & 32.3 & 58.1 & 220.7 & 12.4 & 53.8 & 40.2 & 346.8 & 25.4 & 338.0 & 13.0\\
  Tokens & 986.7 & 1340.1 & 1634.0 & 767.1 & 638.9 & 760.8 & 3634.6 & 469.4 & 1009.0 & 1031.2 & 361.7 & 505.1(7311.4) & 398.9 & 464.6 & N/A\\
  \bottomrule
\end{tabular}
}
\end{table}

\subsection{Future Directions}

Based on our empirical evaluation, we summarize several areas worth future investigation for LLM-powered path constraint solving:

\textbf{Efficient LLM reasoning:} The time efficiency of LLMs, especially LRMs, is substantially lower than that of traditional tools. The majority of LRMs' inference time is spent on the long CoT generated during the reasoning performance. To alleviate this issue, more efficient reasoning strategies can be employed to mitigate the overthinking phenomenon in LRMs. Existing efficient reasoning approaches include both training-based \cite{zhang2024chain} and training-free \cite{yue2024large} methods, and researchers may develop new approaches specially dedicated to the execution path reasoning task.

\textbf{Improving LLM ability via training:} One of the major features of execution path constraint solving is that its results can be accurately verified. For example, in the test case generation task, we can directly measure the correctness of the generated test case by comparing its execution trace with the given input trace, without additional ground-truth labels for evaluation. This creates an opportunity to enhance LLMs on path constraint solving through reinforcement learning from verifiable rewards (RLVR) \cite{guo2025deepseek}. Applying RLVR to smaller LLMs may further help alleviate the reliance on cost-intensive proprietary or large-scale LLMs.

\textbf{LLM-augmented path exploration:} In this study, we mainly focus on the path constraint solving challenge in symbolic execution. Another critical challenge is \textit{path explosion}, where the number of extracted paths becomes too large to analyze them all. The study in RQ2 reveals that LLMs may possess some ability in identifying infeasible paths, a property that could be leveraged to reduce the search space by pruning such paths at an early stage. Another possible solution is to let LLMs prioritize ``important'' execution paths that are most relevant to downstream tasks, such as test case generation or division-by-zero detection.

\subsection{\added{Discussion on Data Leakage}}

\added{As a dataset collected from LeetCode problems and solutions, TestEval has the potential of data leakage. To study whether the data leakage affects the performance of LLMs in test case generation, we conduct a study that split the dataset into two halves by the cutoff date of LLMs. The TestEval dataset is collected from data up to Apr 2024 \cite{wang2025testeval}, which is earlier than the training cutoff date of most LLMs in our study, except o3-mini, whose cutoff date is Oct 2023. Therefore, we choose o3-mini as our analysis object, and evaluate it on TestEval programs that are before or after Oct 2023. We also compare o3-mini with the rest of LRMs, to see whether the cutoff date affects model performances.}

\begin{table}
\captionof{table}{\added{Test case generation results on TestEval data before and after Oct 2023.}}
  \label{tab:cutoff-analysis}
   \scalebox{0.75}{
  \begin{tabular}{lccccccc}
  \toprule
  \added{Model} & \added{P acc before} & \added{P acc after} & \added{Rate (after/before)} & \added{S acc
  before} & \added{S acc after} & \added{B acc before} & \added{B acc after} \\
  \midrule
  \added{o3-mini} & \added{52.8\%} & \added{46.9\%} & \added{0.89} & \added{48.0\%} & \added{31.8\%} & \added{49.2\%}
  & \added{27.1\%} \\
  \midrule
  \added{o4-mini} & \added{67.6\%} & \added{51.6\%} & \added{0.76} & \added{60.3\%} & \added{44.0\%} & \added{61.6\%}
  & \added{44.2\%} \\
  \added{DeepSeek-R1} & \added{65.2\%} & \added{59.4\%} & \added{0.91} & \added{62.6\%} & \added{46.9\%} &
  \added{64.2\%} & \added{46.8\%} \\
  \added{gpt-oss-120b} & \added{55.8\%} & \added{48.4\%} & \added{0.86} & \added{48.6\%} & \added{31.2\%} &
  \added{48.7\%} & \added{25.7\%} \\
  \added{Qwen3-thinking-235b} & \added{66.4\%} & \added{66.7\%} & \added{1.00} & \added{55.9\%} &
  \added{48.5\%} & \added{57.3\%} & \added{42.3\%} \\
  \bottomrule
  \end{tabular}
  }
\end{table}

\added{Table \ref{tab:cutoff-analysis} shows the test case generation metrics before and after Oct 2023. We find that o3-mini, along with 3 other LLMs, demonstrate performance degradation after Oct 2023, the only exception is Qwen3-thinking-235b. We believe that this performance drop is not caused by memorization, but by the difficulty of paths themselves: the average path length before Oct 2023 is 220 statements, while the average path after is 289. On the other hand, 3 LLMs with cutoff date later than Apr 2024 (the cutoff of the complete TestEval dataset) also have a performance drop. When we compute the rate on accuracy after and before, o3-mini achieves a comparable rate (0.89) to those 3 LLMs. The above findings suggest that data leakage is not a primary concern on our test case generation task.}

\subsection{\added{Discussion on Uncertainty}}

\added{As the output of LLMs is not deterministic, there may exist uncertainty for the generated test cases or path classification results in our study. To avoid the uncertainty brought by the randomness of LLMs themselves, we have set the temperature to 0 for all LLMs. However, other factors, such as prompt wording, may also introduce uncertainty. To study the effect of prompts, we generate 3 variant prompt templates for the test case generation task by editing some words without changing the semantics of the prompt. We then run LLMs on all 3 prompt variants, and compute the average accuracies and standard deviation for these variants along with the original prompt templates.}

\begin{center}
\captionof{table}{\added{The results on test case generation with prompt variants.}}
    \label{tab:uncertainty}
    \begin{tabular}{lccc}
    \toprule
    \added{Model} & \added{P acc } & \added{S acc } & \added{B acc } \\
    \midrule
    \added{GPT-4.1} &
    \added{46.3 $\pm$ 1.1\%} &
    \added{46.9 $\pm$ 2.6\%} &
    \added{46.5 $\pm$ 2.7\%} \\
    \added{DeepSeek-v3} &
    \added{34.1 $\pm$ 2.9\%} &
    \added{39.3 $\pm$ 1.4\%} &
    \added{38.7 $\pm$ 1.5\%} \\
    \added{Gemma3-12b} &
    \added{11.7 $\pm$ 0.7\%} &
    \added{18.0 $\pm$ 1.4\%} &
    \added{16.3 $\pm$ 1.3\%} \\
    \added{o4-mini} &
    \added{66.1 $\pm$ 0.4\%} &
    \added{59.7 $\pm$ 2.3\%} &
    \added{59.6 $\pm$ 2.4\%} \\
    \added{GPT-oss-120b} &
    \added{55.8 $\pm$ 1.6\%} &
    \added{43.5 $\pm$ 4.1\%} &
    \added{42.5 $\pm$ 3.9\%} \\
    \bottomrule
    \end{tabular}
\end{center}

\added{Table \ref{tab:uncertainty} demonstrates the results with prompt variants. Among the 5 LLMs involved in this analysis, including both strong proprietary LLMs and small open-source models, all of them exhibit a small variance across different prompt templates. This indicates that LLMs are not sensitive to prompt wording under our test case generation setting, which means the uncertainty is within an acceptable range.}

%% file: sections/threats.tex
The paper has two main threats to validity.

First, LLM augmentation techniques, such as advanced prompting (e.g., in-context learning) and fine-tuning, are not applied in the study. To alleviate this threat, we design prompt templates by following the experience of state-of-the-art software testing benchmarks \cite{jain2025testgeneval}. Moreover, our study only serves as an early-stage exploration of the inherent ability of LLMs to solve path constraints. Future research on advanced prompting and fine-tuning can still be integrated into our evaluation framework.

Second, our study does not involve a direct comparison with traditional symbolic execution tools. As noted earlier, our target language, Python, lacks well-established tool support due to its dynamically typed nature and flexible syntax. Also, existing Python symbolic execution tools cannot support ``solve a given execution path'' task, no matter for test case generation or bug detection. To mitigate this threat, we introduced a comparison of time efficiency between LLMs and the Python symbolic execution tool CrossHair. Considering the limitations of existing tools, the main goal of our study is to explore the possibility of an LLM-powered symbolic execution engine for Python, which may free developers and researchers from the burdens of constructing traditional solver-based tools.